\documentclass[a4paper,fleqn]{cas-dc}

\usepackage[authoryear]{natbib}
\usepackage{mdwlist}
\usepackage[switch]{lineno}

\def\tsc#1{\csdef{#1}{\textsc{\lowercase{#1}}\xspace}}
\tsc{WGM}
\tsc{QE}
\tsc{EP}
\tsc{PMS}
\tsc{BEC}
\tsc{DE}

\begin{document}
\let\WriteBookmarks\relax
\def\floatpagepagefraction{1}
\def\textpagefraction{.001}
\shorttitle{Microporosity and Parent Body of the Rubble-Pile NEA (162173) Ryugu}
\shortauthors{W. Neumann et~al.}

\title [mode = title]{Microporosity and Parent Body of the Rubble-Pile NEA (162173) Ryugu}                      

\author[1,2]{Wladimir Neumann}[type=editor,
auid=000,bioid=1,
orcid=0000-0003-1932-602X]
\cormark[1]
\ead{wladimir.neumann@dlr.de}
\ead[url]{www.researchgate.net/profile/Wladimir_Neumann}

\credit{Conceptualization of this study, Methodology, Software}
\address[1]{Klaus-Tschira-Labor für Kosmochemie, Institut für Geowissenschaften, Universität Heidelberg, Im Neuenheimer Feld 234-236, 69120 Heidelberg, Germany}
\address[2]{Institute of Planetary Research, German Aerospace Center (DLR), Rutherfordstr. 2, 12489 Berlin, Germany}

\author[2]{Matthias Grott}

\author[1]
{Mario Trieloff}

\author[3]
{Ralf Jaumann}
\address[3]{Free University of Berlin, Department of Earth Sciences, Malteserstr. 74-100, 12249 Berlin, Germany}

\author[4]
{Jens Biele}
\address[4]{German Aerospace Center, Microgravity Support Center, Linder Höhe, 51147 Cologne, Germany}

\author[5]{Maximilian Hamm}
\address[5]{University of Potsdam, Institute of Mathematics, Karl-Liebknecht-Str. 24-25, 14476 Potsdam, Germany}

\author[6,7]
{Ekkehard Kührt}
\address[6]{Institute of Optical Sensor Systems, German Aerospace Center, Berlin, Germany}
\address[7]{Qian Xuesen Laboratory of Space Technology, China Academy of Space Technology, Beijing, China}


\begin{abstract}
Both observations of C-type near-Earth asteroids and laboratory investigations of carbonaceous chondritic meteorites provide strong evidence for a high microporosity of C-type asteroids. Boulder microporosity values derived from in-situ measurements at the surface of the rubble-pile NEA (162173) Ryugu are as high as $55$ \%, which is substantially higher than for water-rich carbonaceous chondrite samples and could indicate distinct evolution paths for the parent body of Ryugu and parent bodies of carbonaceous chondrites, despite spectral similarities. In the present study, we calculate the evolution of the temperature and porosity for early solar system's planetesimals in order to constrain the range of parameters that result in microporosities compatible with Ryugu’s high-porosity material and likely burial depths for the boulders observed at the surface. By varying key properties of the parent body, such as accretion time $t_{0}$ and radius $R$ that have strong influence on temperature and porosity and by comparing the interior porosity distribution with the measured boulder microporosity, hydration, and partial dehydration of the material, we constrain a field within the $(R,t_{0})$-diagram appropriate for bodies that are likely to have produced such material. Our calculations indicate a parent body size of only a few km and its early accretion within $\lesssim 2-3$ Myr after the formation of Ca-Al-rich inclusions (CAIs). A gradual final porosity profile of best-fit bodies indicates production of both low- and high-density boulders from the parent body material. By contrast, parent body properties for CI and CM chondrites obtained by fitting carbonate formation data indicate a radius of $\approx 20-25$ km and an accretion time of $\approx 3.75$ Myr after CAIs. These results imply a population of km-sized early accreting highly porous planetesimals as parent bodies of the rubble-pile NEA Ryugu (and, potentially, other NEAs) and a population of larger and late accreting less porous planetesimals as parent bodies of water-rich carbonaceous chondrites.
\end{abstract}

\begin{highlights}
\item We investigate the porosity for ice-rich planetesimals of the early solar system using numerical modeling
\item Our models constrain the conditions which result in parent body porosities that agree with the Ryugu's boulder porosity
\item We constrain Ryugu's parent body radius and accretion time in the range of a few km and $\lesssim 2-3$ Myr after CAIs
\item We examine a potential connection to water-rich carbonaceous chondrites and find a common parent body unlikely
\item Our modeling suggests that Ryugu's parent body belonged to a different planetesimal population than parent bodies of CI and CM chondrites
\end{highlights}


\maketitle

\section{Introduction}

Unveiling key properties of parent bodies of meteorites and rubble pile asteroids is a continuous challenge for planetary science that has implications for the overarching questions of Earth's origin and development of the early solar system, and is, further, of relevance for current planetary defense endeavors. Space missions to near earth objects contribute important findings that can be utilized jointly with results of lab investigations and numerical tools, such as global asteroid evolution models. 

Observations of the C-type near earth asteroid (NEA) (162173) Ryugu by Hayabusa2 demonstrated that this asteroid is a low-density rubble pile \citep{Watanabe2019,Sugita2019,Kitazato2019} whose surface is dominated by large boulders \citep{Michikami2019}. Hayabusa2 payload included the Mobile Asteroid Surface Scout (MASCOT) lander \citep{Ho2017} that was released to the surface of the asteroid and obtained surface brightness temperature measurements for a full day-night cycle using its infrared radiometer MARA \citep{Grott2017}. During MASCOT’s operational phase, measurements provided brightness temperatures of a single boulder that allowed estimating its thermal inertia \citep{Grott2019,Jaumann2019}. Observed thermal properties indicate high boulder microporosities \citep{Grott2019}, consistent with the overall low bulk density of $1190 \pm 20$ kg m$^{-3}$ \citep{Sugita2019} and a high bulk porosity of $\approx 50$ \% \citep{Watanabe2019} derived by assuming a grain density of a typical carbonaceous chondrite \citep{Britt2002,Macke2011,Flynn2018}. Boulder thermal properties imply low thermal conductivity values of $k = 0.06-0.16$ Wm$^{-1}$K$^{-1}$ and high boulder microporosity of $\phi_{boulder}\approx 28-55$ \% for different models of porosity-dependent thermal conductivity $k(\phi)$ \citep{Flynn2018,Henke2016}.

The bulk porosity of a rubble pile asteroid is derived from the contributions of the microporosity (the intrinsic porosity of boulders) and the macroporosity (the voids in-between boulders), and, in its entirety, it is a result of processes that occurred during the thermal evolution of the parent bodies, during their destruction, and during the re-accretion of the rubble that eventually formed Ryugu, with subsequent arrangement of pieces and a potential contribution by post-re-accretional evolution. By contrast, the porosity of a single boulder is a local microporosity value resulting mainly from processes that took place during the thermal evolution of a parent body from which the boulder originated prior to their re-accretion onto Ryugu. Based on the observed size-frequency distribution of boulders on the surface, \citet{Grott2020} found through the application of a mixing model to the boulder size distribution a macroporosity of $16 \pm 3$ \% and derived an average grain density of $2848 \pm 152$ kg m$^{-3}$ for a boulder microporosity of $50$ \%, where the grain density is consistent with values obtained for CM and the Tagish Lake meteorites.

Evidence for the boulder composition comes from both pre-flight and in-orbit spectroscopic observations that showed presence of phyllosilicates at the surface and general consistency with partially hydrated carbonaceous meteorites, in particular, with moderately dehydrated CI and CM chondrites \citep{Moskovitz2013,Perna2017,Sugita2019}. This implies that the boulder material was produced from a water-rich parent body that accreted from dust and ice and argues for water action at least in the form of hydration of dry primordial minerals (while evidence for water flow and differentiation of the parent body is lacking). It also indicates an origin beyond the frost line, where water ice could condense in the protoplanetary disk when first planetesimals formed. A large initial heliocentric distance implies, further, relatively cold initial conditions with an initial planetesimal temperature of not more than $\approx 170$ K. This is close to typical values representative for CI or CM parent bodies and outer belt asteroids accreted at $\approx 2.7$ AU \citep[e.g.,][]{Hayashi1981,Wakita2011,Bland2017} or even close to the H$_2$O condensation temperature of $\leq 150$ K under typical solar nebula conditions.

Spectral similarity with carbonaceous chondrites brings about the possibility that parent body properties derived for these meteorites could also be valid for Ryugu's parent body. Notably aqueously altered carbonaceous chondrites that are abundant in hydrated minerals are CI, CM, and CR groups. Their approximate alteration temperatures range from $273$ K to $403$ K for CM \citep{Guo2007,Alexander2015,Fujiya2012}, from $<323$ K to $423$ K for CI \citep{Leshin1997,Clayton1984,Clayton1999,Zolensky1993,Fujiya2013}, and $<423$ K \citep{Zolensky1993}, although medium-temperature alteration within $323-570$ K was suggested recently for CM chondrites as well \citep{Verdier2017}. Thereby, the petrographic types range up to $2$ for CI and CM and up to $3$ for CR chondrites, lacking significant thermal metamorphism. Of note are the Yamato-type (CY) carbonaceous chondrites that are similar to CI and CM groups, but petrographically distinct from those. They indicate decomposition and metamorphism of aqueously altered Mg-Fe-rich carbonates at $773-1073$ K and were suggested to be derived from a near-Earth source based on their short cosmic-ray exposure ages. These constraints on the thermal evolution are not stringent. While experimental studies on the CV chondrite Allende showed that aqueous alteration of its minerals requires weeks at $> 420$ K \citep{Jones2006}, the alteration time scale increases only moderately on geological timescales for lower temperatures, requiring, for instance, approximately $200$ years at $T = 273$ K for serpentinization \citep{Neumann2020a}.

The sequence of events that preceded the formation of Ryugu comprises at least accretion of an original parent body, its disruption, and re-accretion of Ryugu as a rubble pile. A more detailed sequence consisting of an original parent body evolution, potentially a series of disruption events and accretion of intermediate parent bodies, and the final accretion to Ryugu was suggested by \citet{Sugita2019}. In this scenario, the original parent body forms early with abundant $^{26}$Al as a source of internal heating and both hydrates and dehydrates later due to internal heating prior to its disruption, as suggested by a comparison between remote-sensing data and meteoritic samples, as well as by the general color uniformity across the surface. It is obvious to conclude that such a scenario favors an onion shell structure of the original parent body, with an increasing degree of thermal metamorphism with depth, and it can be investigated quite well with thermal evolution models for planetesimals heated by radioactive decay.

Estimates of material accumulation from a $100$-km diameter parent body after a catastrophic disruption \citep{Sugita2019} indicate that materials from all depths contribute to each small object produced, consistent with the spectral homogeneity of the asteroid and a limited local heterogeneity of boulders observed by Hayabusa2. This would necessitate a homogeneous temperature structure of the interior of the parent body except in a thin surface layer and nearly homogeneous material properties at any depth, such as petrographic type, mineralogical composition, and porosity. However, the interior of moderately heated less than few tens of km sized planetesimals does not develop homogeneously, in particular with regard to the porosity \citep{Neumann2014}, implying a certain depth range as source region for Ryugu, if the parent body was small. Furthermore, small parent bodies of up to $40$ km radius were expectably more numerous in the early Solar System than large ones with radii of $\gtrsim 100$ km, and hence more probable as the source of rubble-pile NEAs.

To determine whether Ryugu's parent body is similar to the parent bodies of CI and CM chondrites, characteristic properties of the latter can be derived from the analysis of formation of various mineralogical components, for example, carbonates (calcite, dolomite, breunnerite) that are secondary minerals formed in the presence of aqueous solutions. While a detailed analysis is out of scope of the present study, we note that overall carbonate formation ages in CI and CM meteorites of $4563.1 - 4561.8$ Myr (corresponding to a formation time of $4.8 - 6.1$ Myr after CAIs \citet{Jilly2017}) are quite similar and provide, along with the precipitation temperature range of $\approx 293-423$ K data points that need to be approximated by the evolution of the temperature in different regions of the parent body in the respective time interval. However, a striking difference between the upper value of $\phi_{boulder}$ and class average porosities of CI and CM chondrites of $34.9$ \% and $22.2$ \%, respectively \citep{Flynn2018}, would require an explanation, such as break-up of high-porosity samples during the atmospheric entry \citep{Grott2019}, even if the above data can be fitted for Ryugu's precursor.

In the present study, we investigate the microporosity of early solar system planetesimals using global thermal evolution and compaction models for the porosity of two-component mixtures of spherically symmetric bodies and reproduce the microporosity derived for the boulders observed by Hayabusa2 in the interior of these bodies. Such models predict the microporosity of planetesimal material established due to internal heating within creep processes that are driven by the joint action of temperature and pressure under conditions that favor creep of constituent materials. No notable creep processes can be expected in a late formed and small object such as Ryugu or in an intermediate object between the first and last disruption events. Furthermore, the reassembly of the material after a catastrophic disruption influences the macroporosity of the asteroid as a whole, but not the microporosity of its constituent boulders. Therefore, assuming that microporosity changes little after the disruption of the parent body, we calculate the microporosity throughout the interior of planetesimals with different sizes and accretion times in order to reproduce the boulder microporosity. We identify potential parent bodies for Ryugu’s material and likely burial depths within these bodies for the boulders observed at the surface of the NEA. By varying key properties of the parent body, such as accretion time $t_{0}$ and radius $R$ that have strong influence on temperature and porosity and by comparing the interior porosity distribution with the measured boulder microporosity, hydration, and partial dehydration of the material, we constrain a field within the $(R,t_{0})$-diagram appropriate for bodies that are likely to have produced such material. By fitting the temperature and age data for the formation of secondary minerals in CI and CM chondrites, we obtain the properties of their parent bodies and compare them with our best-fit results for the parent body of Ryugu, examining in this manner a potential connection suggested by Ryugu's spectral properties.

\section{Methods}
From the assumption that rubble pile NEA material is a product of the early solar system planetesimal population, a numerical model for the evolution of such bodies was used for calculations. The 1D finite differences thermal evolution model for $^{26}$Al-heated water-rich planetesimals was built on the basis of those presented in \citet{Neumann2015}, \citet{Neumann2019a}, and \citet{Neumann2020a}. It considers heating of small bodies after accretion as porous aggregates and the evolution of the temperature in their interiors as well as compaction of hydrated material from an initially unconsolidated state due to hot pressing by solving a number of equations that describe these processes. In the following, the model features specific to this study are outlined briefly, while for details we refer to the above publications.

The basic equation is the non-stationary 1D heat conduction equation in spherical coordinates that is discretized by the finite differences method along the spatial and temporal domain and solved for the temperature, with the radiogenic decay being the energy source for the temperature change. Using typical data for aqueously altered carbonaceous chondrites \citep[e.g.,][]{Neumann2020a}, we include both short-lived radionuclides $^{26}$Al and $^{60}$Fe and long-lived radionuclides $^{40}$K, $^{232}$Th, $^{235}$U, and $^{238}$U as heat sources. We assume that the radionuclides are homogeneously distributed within the material. However, the heat source density scales with the porosity $\phi$ that is constant throughout the interior upon the accretion and develops inhomogeneously during the thermal evolution. No further heat sources are invovled.

A water-rich composition of Ryugu's precursor body or bodies evidenced by observed traces for aqueous alteration imposes the necessity of considering multiple effects that arise from such a composition. Here, we include consumption of the latent heat of ice melting, influence of the water on the porosity evolution, presence of hydrous minerals upon aqueous alteration, and material properties.

In agreement with spectral observations of Ryugu that suggest a composition close to CI or CM chondrites \citep{Moskovitz2013,Perna2017,Sugita2019}, we assume properties corresponding to an ice-rich initial composition that leads to a material dominated by phyllosilicates upon aqueous alteration with $\gtrsim 84$ vol.\% antigorite serpentine and $\lesssim 16$ vol.\% olivine. This is a rough representation of the composition and does not contain minor species, such as various sulfates, sulfides, or carbonates. It is, however, reasonably representative in terms of the thermal evolution and compaction behavior and not far from typical CI and CM compositions in the terms of mineralogy and chemistry \citep{Howard2011,King2015}. Notably, representatives of both CI and CM chondrites contain roughly around $25$ wt.\% of iron, which is present mostly in the form of silicates or oxides. An implication of the calculations for the highly energetic thermal conditions in the planetesimals accreting close to the formation of the CAIs would be the formation of a metallic core made of this iron fraction \citep{Neumann2018b} after potential reduction of iron oxides to metallic iron. Since such an evolution path is unlikely for Ryugu, iron oxide reduction and core formation are not modeled here.

A pre-hydration ice mass fraction of $0.2$ is used in agreement with the final hydration level of a CM chondrite with approximately $1:1$ hydrous/anhydrous mineral ratio \citep{Brearley2006}. The latent heat of $3.34\cdot 10^{5}$ J kg$^{-1}$ is consumed in a temperature interval of two degrees around the melting temperature of $T = 273$ K in order to avoid numerical issues with too sharp a phase transition at $273$ K and contributes to the energy balance via the modification of the heat capacity with a Stefan number weighted with the above ice mass fraction \citep[e.g.,][]{Neumann2012}.

It is important to note that the present study does not consider water-rock differentiation since in this specific setup only a negligible free water fraction remains after most water is consumed for hydration of dry silicates. Therefore, further processes resulting from extensive water flow, such as hydrothermal convection, compaction of a rocky core with interstitial water, convection of a water layer, or solid-state convection of an ice crust do not apply here.

We use a typical approach to model the compaction of planetesimals by hot pressing adopted in several studies of rocky or icy small bodies. For more details, we refer to \citet{Neumann2014}, where compaction of small planetesimals was explored and to \citet{Neumann2020a}, where it was modeled in an analogous manner while considering differentiation of water-rich planetesimals and the dwarf planet Ceres. The variation of the bulk pore space volume fraction (termed as porosity $\phi$) is described by time-dependent differential equations which establish the relation between the strain rate $\dot{\varepsilon}$ and the applied ("effective") stress $\sigma$. For the mineralogical composition considered, the volume fractions $v_{i}$ and the associated porosities $\phi_{i}$ correspond to (antigorite) serpentine ($i=se$) or olivine ($i=ol$). The creep laws for the single mineral phases provide the strain rates $\dot{\varepsilon}_{i}$.
A Peierl’s law equation \citep{Katayama2008} derived for the deformation of antigorite at pressures of $\leq 200$ MPa by \citet{Amiguet2012}
\begin{eqnarray}
    \dot \varepsilon_{se} &=& \frac{\partial \log ( 1-\phi_{se} )}{\partial t} \nonumber \\
    &=& 4\cdot 10^{-22} \sigma^{2} \exp\left(-\frac{27}{\mathcal{R}T} \left( 1-\frac{\sigma}{2.7\cdot 10^{9}} \right) \right) \label{cr1}
\end{eqnarray}
and a diffusion creep law for olivine derived by \citet{Schwenn1978}
\begin{eqnarray}
\dot \varepsilon_{ol} &=& \frac{\partial \log ( 1-\phi_{ol} )}{\partial t} \nonumber \\
&=& 1.26\cdot 10^{-18} \sigma^{1.5} b^{-1.4} \exp \left(-\frac{356}{\mathcal{R}T} \right) \label{cr2}
\end{eqnarray}
were used for the calculation of the porosity. Note that the stress $\sigma$ is in Pa, the grain size $b$ in m, the activation energy $\mathcal{E}$ in kJ mol$^{-1}$, the gas constant $\mathcal{R}$ in kJ and the temperature $T$ in K. An enhancement of the strain rate in presence of fluids is accounted for by multiplying the right-hand side of equations (\ref{cr1}) and (\ref{cr2}) with the term $\exp(\alpha \chi_{\text{melt}})\exp(\alpha \chi_{\text{water}})$ \citep{Mei2002}, where $\alpha = 25$, $\chi_{\text{melt}}$ is the cumulate iron and silicate melt volume fraction (assuming linear melting between $1262$ K and $1689$ K for iron and $1440$ K and $1736$ K for silicates), and $\chi_{\text{water}}=0.1$ is the volume fraction of free water assumed after the melting of ice in the initial ice-dust mixture. From the volume fraction weighted arithmetic mean of strain rates of both species, the average local strain rate
\begin{eqnarray}
\dot{\varepsilon}=\frac{\partial \log (1-\phi)}{\partial t}= v_{se}\dot{\varepsilon}_{se}+v_{ol}\dot{\varepsilon}_{ol}
\end{eqnarray}
and the average local porosity $\phi$ are obtained.

\begin{table*}
\centering
\caption{Parameters used in the models. The data for the calculation of the radiogenic energy production are from \citet{Barrat2012}, \citet{Kita2013}, \citet{Tang2012}, \citet{Finocchi1997}, \citet{VanSchmus1995}. The element mass fractions refer to stable isotopes, the initial ratios are between unstable and stable isotopes of an element, and the decay energies are per particle.}
\centering
\begin{tabular}{lrrrc}
\hline \\ [-1.7ex]
Variable & Symbol & Unit & Value & Reference \\
\\ [-1.7ex]
\hline
\\ [-1.4ex]
Initial porosity & $\phi_{0}$ & - & $0.5$, $0.6$, $0.7$, $0.8$ & \\
 \\ [-1.4ex]
Initial H$_{2}$O mass fraction & $x_{\text{H}_{2}\text{O,0}}$ & - & $0.2$ & \citet{Brearley2006} \\
\\ [-1.4ex]
Effective stress & $\sigma$ & Pa & \multicolumn{2}{c}{see text and \citet{Neumann2014}} \\
\\ [-1.4ex]
Grain size & $b$ & m & $1\cdot 10^{-6}$ & \citet{Cloutis2011a,Cloutis2011b} \\
\\ [-1.4ex]
Gas constant & $\mathcal{R}$ & J mol$^{-1}$K$^{-1}$ & $8.314472$ & - \\
\\ [-1.4ex]
Ambient temperature & $T_{\text{S}}$ & K & $170$ & \citet{Bland2017} \\
\\ [-1.4ex]
Water latent heat & $L_{\text{H}_{2}\text{O}}$ & J kg$^{-1}$K$^{-1}$ & $3.34 \cdot 10^{5}$ & - \\
\\ [-1.4ex]
Stefan-Boltzmann constant & $\sigma_{\text{SB}}$ & W m$^{-1}$K$^{-1}$ & $5.67 \cdot 10^{-8}$ & - \\
\\ [-1.4ex]
Gravitational constant & $G$ & m$^{3}$kg$^{-1}$s$^{-2}$ & $6.67 \cdot 10^{-11}$ & - \\
\\ [-1.4ex]
Serpentine th. conductivity & $k_{se}$ & W m$^{-1}$K$^{-1}$ & $(0.404+2.46 \cdot 10^{-4} T )^{-1}$ & \citet{Grindrod2008} \\
\\ [-1.4ex]
Olivine th. conductivity & $k_{ol}$ & W m$^{-1}$K$^{-1}$ & $4.3$ & \citet{Horai1972} \\
\\ [-1.4ex]
Chondrite heat capacity & $c_{p}$ & J kg$^{-1}$K$^{-1}$ & $800+0.25T-1.5\cdot 10^{7}T^{-2}$ & \citet{Yomogida1984} \\
\\ [-1.4ex]
Grain density & $\rho_{g}$ & kg m$^{-3}$ & $2670$ & \citet{Grott2019b} \\
 \\ [-1.4ex]
Serpentine grain density & $\rho_{se}$ & kg m$^{-3}$ & $2500$ & \\
\\ [-1.4ex]
Olivine grain density & $\rho_{ol}$ & kg m$^{-3}$ & $3580$ & \\
\\ [-1.4ex]
Serpentine volume fraction & $v_{se}$ & - & $0.84$ & \\
\\ [-1.4ex]
Olivine volume fraction & $v_{ol}$ & - & $0.16$ & \\
[0.6ex]
\end{tabular}
\begin{tabular}{c|llllll}
\hline \\ [-1.7ex]
Isotope & \multicolumn{1}{c}{$^{26}$Al} & \multicolumn{1}{c}{$^{60}$Fe} & \multicolumn{1}{c}{$^{40}$K} & \multicolumn{1}{c}{$^{232}$Th} & \multicolumn{1}{c}{$^{235}$U} & \multicolumn{1}{c}{$^{238}$U} \\
\\ [-1.7ex]
\hline
\\ [-1.4ex]
Element mass fraction & $1.18\cdot 10^{-2}$	& $2.12\cdot 10^{-1}$ &	$3.42\cdot 10^{-4}$	& $3.98\cdot 10^{-8}$ & $1.12\cdot 10^{-8}$ & $1.12\cdot 10^{-8}$ \\
\\ [-1.4ex]
Half-life [years] & $7.17\cdot 10^{5}$ & $2.62\cdot 10^{6}$ & $1.25\cdot 10^{9}$ & $1.41\cdot 10^{10}$ & $7.04\cdot 10^{8}$ & $4.470\cdot 10^{9}$ \\
\\ [-1.4ex]
Initial ratio & $5.25\cdot 10^{-5}$ & $1.15\cdot 10^{-8}$ & $1.50\cdot 10^{-3}$ & $1.0$ & $0.24$ & $0.76$ \\
\\ [-1.4ex]
Decay energy [J] & $4.99\cdot 10^{-13}$ & $4.34\cdot 10^{-13}$ & $1.11\cdot 10^{-13}$ & $6.47\cdot 10^{-12}$ & $7.11\cdot 10^{-12}$ & $7.610 \cdot 10^{-12}$ \\
[0.6ex]
\hline
\end{tabular}
\label{table1}
\end{table*}

In the model, we use this approach for the calculation of the evolution of the dust porosity (i.e., we define $\phi_{0}$, $\phi$ and $\phi_{i}$ as well as $\dot{\varepsilon}$ and $\dot{\varepsilon}_{i}$ for the calculation of creep while $\phi>0$). No melt porosity after the melting of water ice is considered. Important free parameters are grain size $b$ and the initial porosity $\phi_{0}$. A grain size of $b = 1$ $\mu$m is based on the matrix grain sizes of CI and CM meteorites \citep{Cloutis2011a,Cloutis2011b}. An initial porosity of $50$ \% is used typically in compaction models of planetesimals and is based on the porosities of the random loose and random close packings as well as on the porosity of $40-50$ \% expected after “cold pressing” acted on a granular material \citep[e.g.,][]{Henke2012,Neumann2014}. Based on a high porosity estimate for the boulders on Ryugu \citep{Grott2019} and on a more porous precursor than the boulders, we consider initial porosities between $50$ \% and $80$ \%, where the upper bound is supported by DEM-based dynamic simulations for $1$ $\mu$m fine particles \citep{Yang2000} and a range of $60$ \% to $80$ \% is supported by the nucleus bulk porosity estimates for the comet 67P \citep{Paetzold2019}. Another essential parameter is the effective stress $\sigma$, which is calculated for some specific packing of equally sized spheres \citep{Neumann2014}. Here, the simple cubic packing that allows for the largest initial porosity of $\approx 50$ \% is used. However, in this geometric model $\sigma$ is not defined beyond this value, although initial porosities adopted surpass it. Therefore, we assume $\sigma=10^{2}P$ outside of its domain of definition and adopt a cutoff for the initial change of the porosity via $\sigma=min\lbrace 10^{2}P,\sigma_{\text{eff}} \rbrace$, where $\sigma_{\text{eff}}$ abides by the definition from the geometric model and $P$ is the lithostatic pressure.

Porosity evolution in icy bodies has been modeled using a similar equation that has a different shape resulting in a different porosity behavior near zero \citep[e.g.,][]{Bierson2018}. This model is based on a two-phase flow setup and on the physical observation of closure of a water-filled borehole in an ice layer \citep{Fowler1984}. Assuming equivalence of surface and volume averaging, an equation for the change of the average cylindrical borehole volume is derived, where the porosity is equivalent to the average borehole radius \citep{Fowler1984}. In addition, the stress is equivalent to the pressure contrast between the water pressure and the lithostatic pressure. By contrast, our model is based on the deformation of a powder compact in a die and on hot pressing experiments \citep[see ][for details]{Neumann2014}. No assumption on the powder compact pore geometry is involved. In addition, a microscopic geometry assumption is used for the calculation of the effective stress, that varies by several orders of magnitude with the size of sintering necks between dust particles and approaches the lithostatic pressure as the porosity converges to zero. Our approach is more suited for calculating porosity evolution for bodies that are not dominated by ice and where water is consumed completely during hydration. In addition, it is less specific in the terms of geometry and of composition by contrast to models based on \citet{Fowler1984}, while the latter are likely more suited for questions related to the ice crust evolution of icy moons.

The material properties, such as bulk values of the density $\rho_{bulk}$, thermal conductivity $k_{b}$, and specific heat capacity $c_p$, vary with the porosity or temperature. The local bulk density scales in the model with the local porosity as $\rho_{bulk}=(1-\phi)\rho_{g}$. The grain density of the compacted material with $\phi = 0$, $\rho_{g}=2670$ kg m$^{-3}$, agrees with that derived by \citet{Grott2019b} from in-situ measurements and boulder size distribution mixing model. This value also fits the composition used here via the equation $\rho_{g}=v_{se}\rho_{se}+v_{ol}\rho_{ol}$ with volume fractions and grain densities of serpentine and olivine (Table $\ref{table1}$), and is bracketed by the average grain densities of CI and CM chondrites of $\approx 2400$ kg m$^{-3}$ and $\approx 3000$ kg m$^{-3}$, respectively \citep{Consolmagno2008,Macke2011}. For the heat capacity, we use the analytic approximation from \citet{Yomogida1984} for a non-differentiated bulk chondritic material. The bulk thermal conductivity $k_{b}$ varies with porosity as $k_{b}=k f(\phi)$, with $k = k_{se}^{v_{se}}k_{ol}^{v_{ol}}$, where the thermal conductivities and volume fractions of serpentine and olivine are involved (Table \ref{table1}). The function $f(\phi)$ varies between $\mathcal{O}(0.001)$ and $1$ if the dust porosity varies between its high initial value of $\phi_0$ and zero \citep{Henke2016}:
\begin{eqnarray}
f(\phi)=\left( \max \lbrace 1-2.216\phi , 0 \rbrace ^{4}+\exp \left( -1.2-\phi/0.167 \right)^{4} \right)^{1/4}\text{.}
\end{eqnarray}
In addition, $k_{eff}=0.089 k_{b}Ra^{1/3}$ \citep[with the Rayleigh number $Ra$, e.g.,][]{Neumann2018b} is used instead of $k_{b}$ for a degree of melting of $\geq 50$ \% defined by a temperature of $1650$ K, i.e., in a mixed iron-silicate magma ocean, without considering any sort of differentiation of Fe-FeS or silicate melts. The latter can, in principle, occur in a certain range of parameter value combinations, with respect to parent body accretion time and size, but will not occur for best-fit parent bodies.

The size of the body itself, defined here by its radius $\overline{R}(t)$ changes with the bulk porosity $\phi_{bulk}(t)$ at the time $t$ according to 
\begin{eqnarray}
\overline{R}(t)=(1-\phi_{bulk}(t))^{-1/3}R \text{,} \label{radius}
\end{eqnarray}
where $R$ is the reference radius, i.e., the radius that would be attained if the porosity were zero. While $\overline{R}(t)>R$ is always true and $R$ is never attained in the calculations, it is used for the analysis of the results.

All equations involved are solved on the domain from the center of the planetesimal up to its surface. The spatial grid is transformed from $0\leq r\leq R$, with the distance from the center $r$ in m, to $0\leq \eta \leq 1$ according to the transformation $\eta :=r/\overline{R}(t)$. The time and space derivatives change as well \citep{Merk2002} and are applied to all equations involved, such that features like Langrangian transport of porosity \citep[][Eq. (10)]{Neumann2012} and other quantities are accounted for. While the positions of the grid points between $0$ and $1$ are fixed, the variable values at the grid points are updated at every time step according to the above transformations. Non-stationary equations are discretized also with respect to the time variable $t$ and solved using implicit finite difference method.

The criteria for a planetesimal to be accepted as Ryugu's parent body candidate applied here are (a) production of a relatively high amount of material with $\phi\approx\phi_{boulder}$; (b) aqueous alteration of a reasonable fraction of such material; (c) agreement of the temperature evolution of such material with partial dehydration of phyllosilicates. In addition, relationship to the CI or CM parent bodies can be examined by (d) fits of the thermal evolution to the carbonate formation ages and temperatures in CI and CM chondrites; (e) comparison of the porosity at the depths of the temperature fits with those of CI and CM meteorites.

\section{Results}

\begin{figure*}
\begin{minipage}[ht]{8.5cm}
\setlength{\fboxsep}{0mm}
\centerline{\includegraphics[trim = 10mm 0mm 10mm 0mm, width=8.5cm]{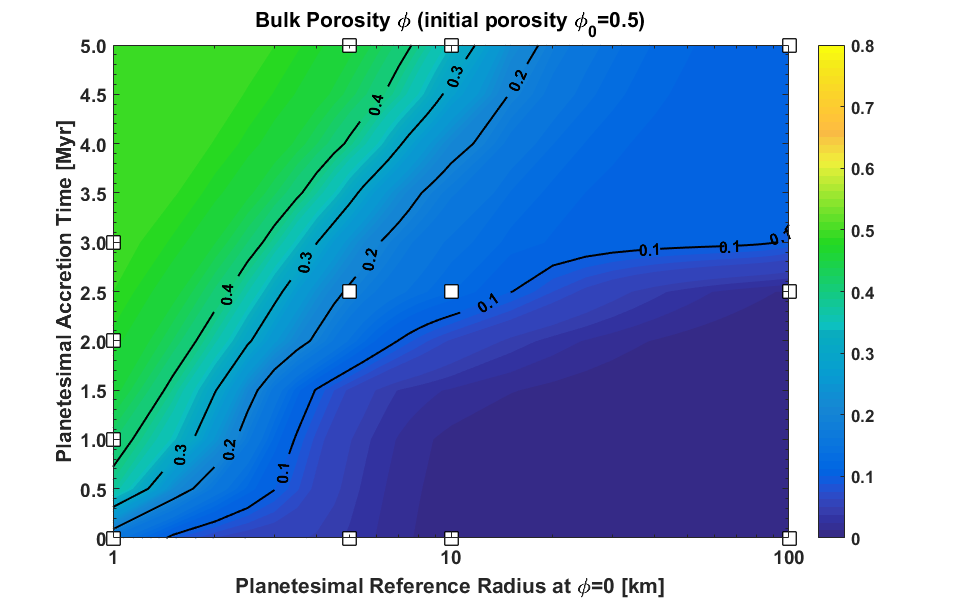}}
\end{minipage}
\begin{minipage}[ht]{8.5cm}
\setlength{\fboxsep}{0mm}
\centerline{\includegraphics[trim = 10mm 0mm 10mm 0mm, width=8.5cm]{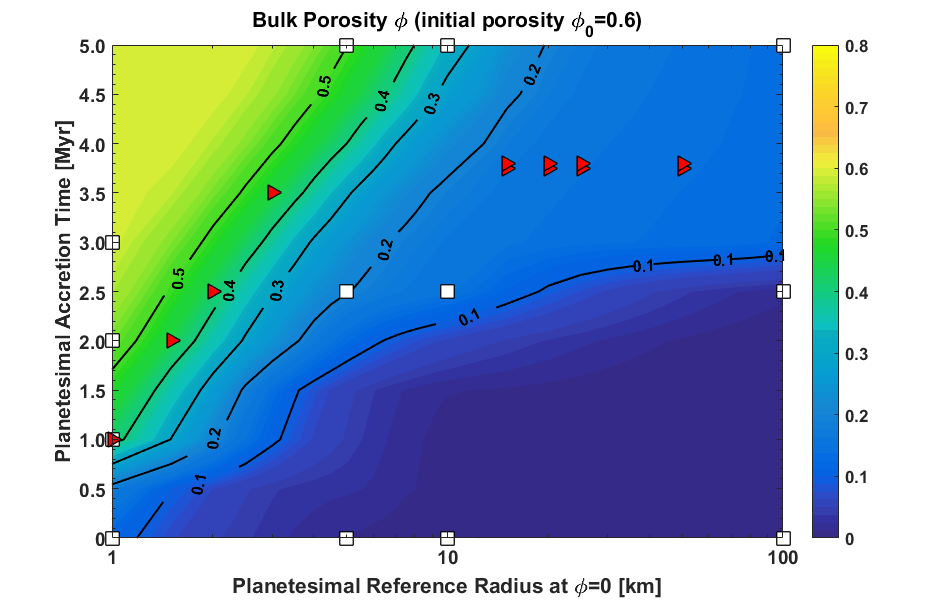}}
\end{minipage}
\\
\begin{minipage}[ht]{8.5cm}
\setlength{\fboxsep}{0mm}
\centerline{\includegraphics[trim = 10mm 0mm 10mm 0mm, width=8.5cm]{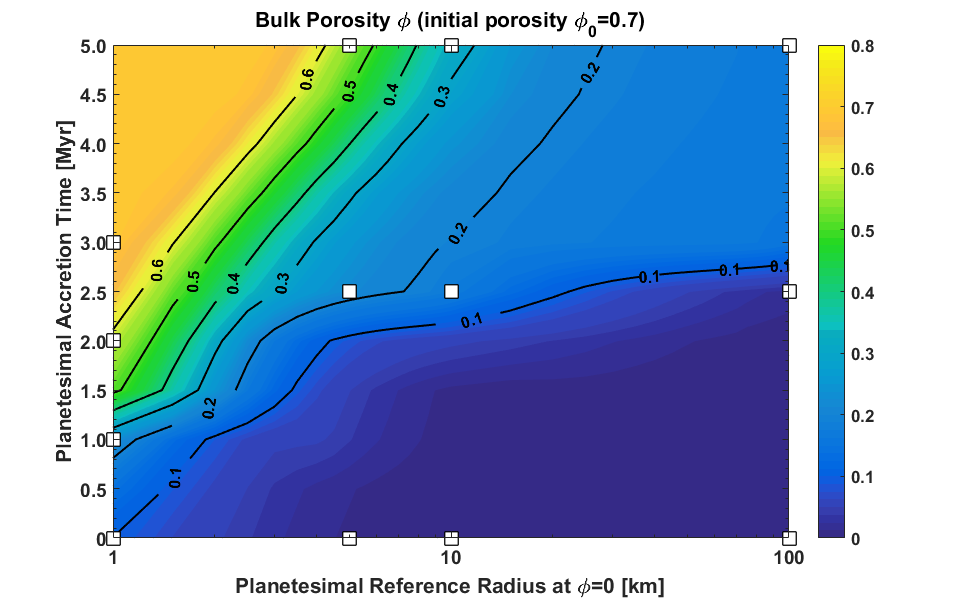}}
\end{minipage}
\begin{minipage}[ht]{8.5cm}
\setlength{\fboxsep}{0mm}
\centerline{\includegraphics[trim = 10mm 0mm 10mm 0mm, width=8.5cm]{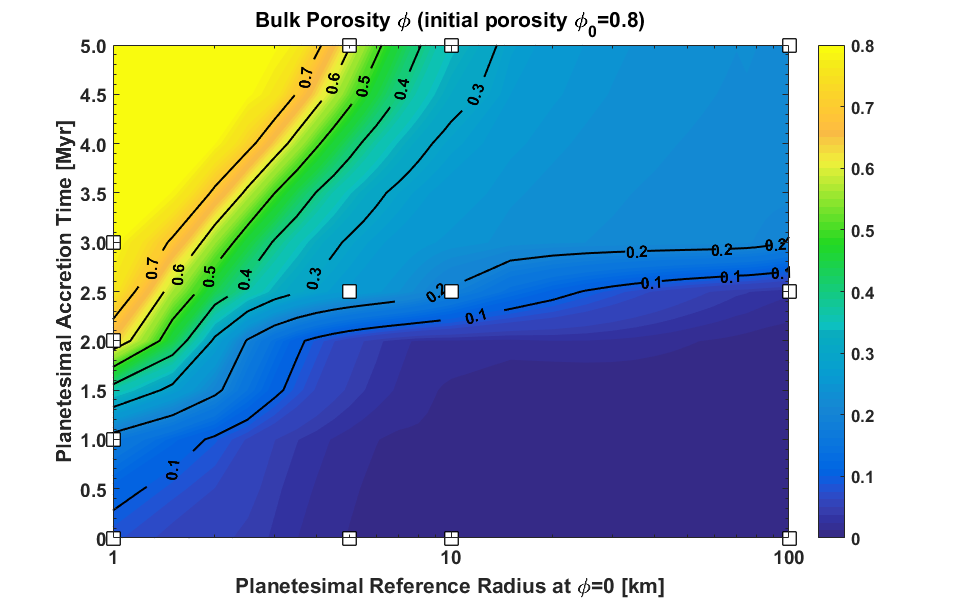}}
\end{minipage}
\caption{The bulk porosity of a planetesimal established after the initial heating phase due to $^{26}$Al and the cessation of compaction due to the cooling as a function of the planetesimal reference radius (i.e., the radius of a body with an equal intrinsic volume but zero porosity; horizontal axes) and accretion time relative to the formation of CAIs (vertical axes). The panels correspond to different assumptions on the initial porosity $\phi_0$ of $50$ \% (top left), $60$ \% (top right), $70$ \% (bottom left), and $80$ \% (bottom right). White squares correspond to bodies addressed in Fig. \ref{fig2} while red triangles correspond to bodies shown in Fig. \ref{fig4}.}
\label{fig1}
\end{figure*}

A rubble pile object is, by definition, a product of one or multiple break-ups of one or several parent bodies. Any metamorphism or alteration of the rubble pile material occurred partially if not totally on a parent body with poorly constrained properties. A minimum requirement on the mass of Ryugu's parent body is the mass of Ryugu itself and a minimum requirement on its initial porosity $\phi_{0}$ is the boulder microporosity of $\phi_{boulder}\approx 28-55$ \%. Therefore, the evolution of precursors that were at least as massive must be calculated. Thereby, any initial porosity $\phi_{0}>0$ implies a larger initial radius than the reference radius for an equal mass (Eq. (\ref{radius})) and, potentially, shrinking due to compaction. The circumstances of the parent body accretion, i.e., its timing and rate, are estimated by our current understanding of the planetesimal formation in the early solar system, spanning, roughly speaking, the first few million years after the formation of CAIs. Apart from that, the parent body mass/size, the initial porosity, and the accretion time are free parameters. For the calculations presented in the following, the initial porosity $\phi_{0}$ values of $0.5$, $0.6$, $0.7$, and $0.8$ were used. For each $\phi_{0}$ value, $(R,t_{0})$ pairs were considered, where the reference radius $R$ varied between $1$ km and $100$ km, while the accretion time $t_{0}$ varied between $0$ Myr and $5$ Myr after CAIs' formation. Adopting these bounds, model runs were performed for each $(R,t_{0})$ pair, where the evolution of the temperature and porosity of initially highly porous primordial planetesimals was calculated.

\begin{figure*}
\begin{minipage}[ht]{8.5cm}
\centerline{\fbox{\includegraphics[trim = 5mm 20mm 10mm 5mm, width=8.3cm]{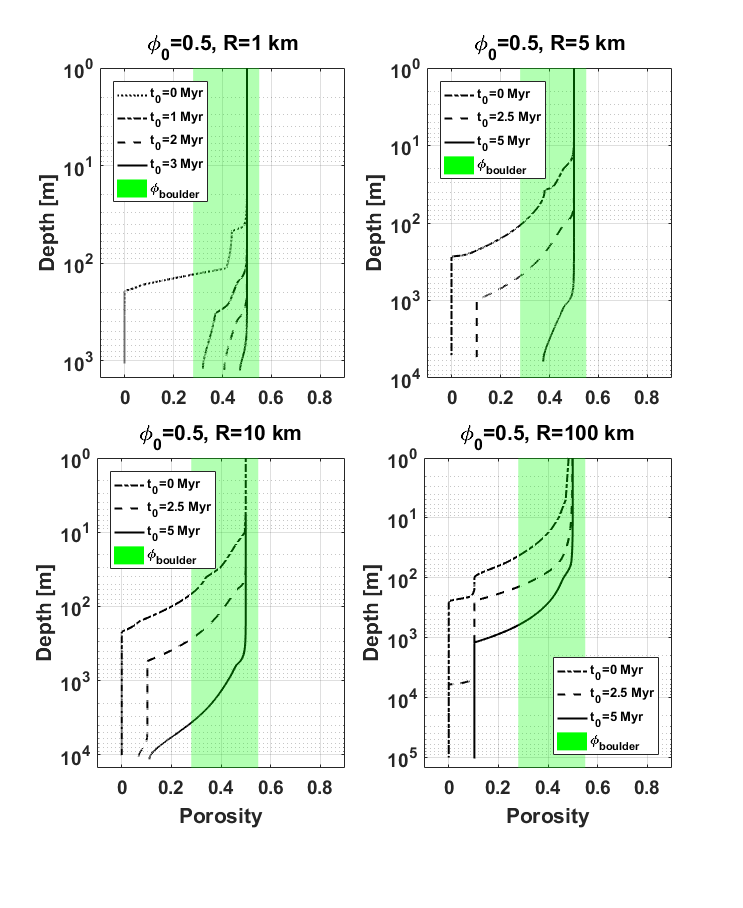}}}
\end{minipage}
\hfill
\begin{minipage}[ht]{8.5cm}
\centerline{\fbox{\includegraphics[trim = 5mm 20mm 10mm 5mm, width=8.3cm]{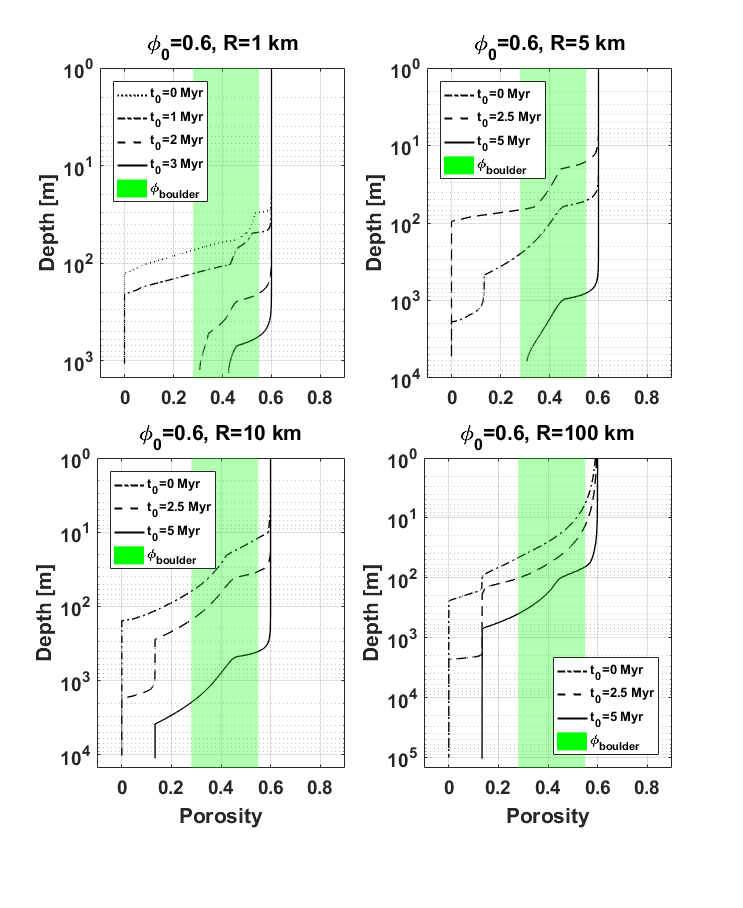}}}
\end{minipage}
\\
\begin{minipage}[ht]{8.5cm}
\centerline{\fbox{\includegraphics[trim = 5mm 20mm 10mm 5mm, width=8.3cm]{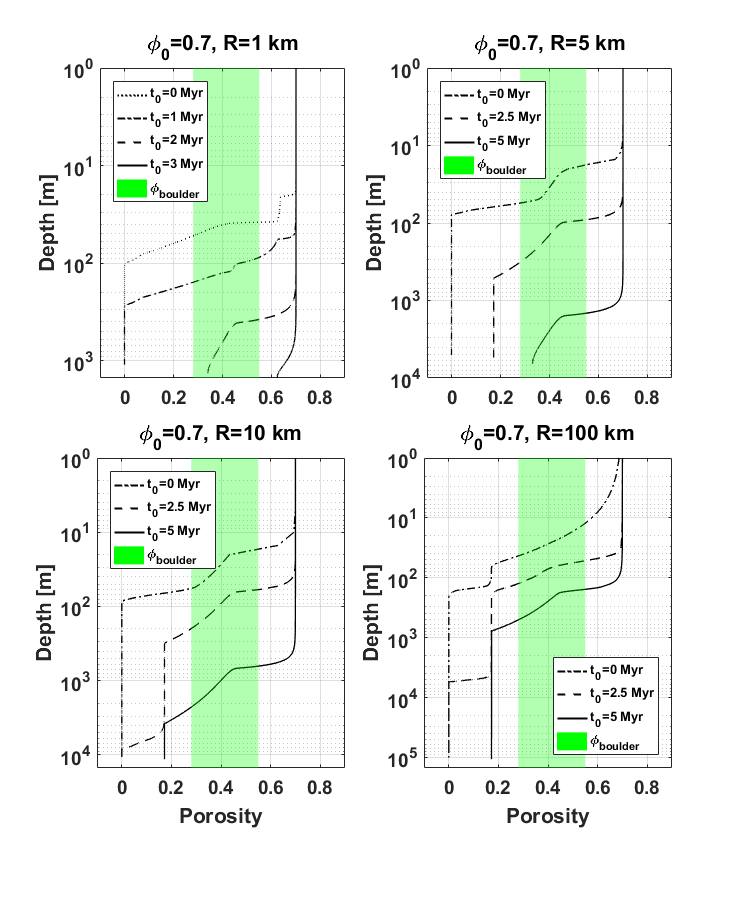}}}
\end{minipage}
\hfill
\begin{minipage}[ht]{8.5cm}
\centerline{\fbox{\includegraphics[trim = 5mm 20mm 10mm 5mm, width=8.3cm]{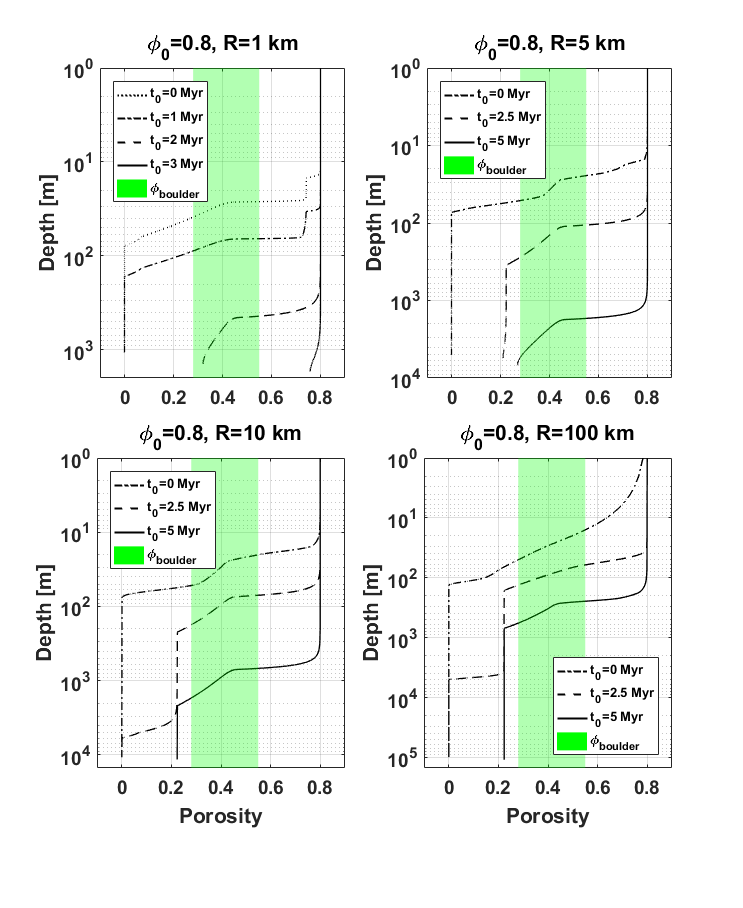}}}
\end{minipage}
\caption{Final porosity profiles for reference radii of $1$, $5$, $10$, and $100$ km (left to right and top to bottom within each box). The boxes corresponds to the initial porosities of $0.5$, $0.6$, $0.7$, and $0.8$ (left to right and top to bottom). For each single panel, porosity profiles for different accretion times are shown and compared with the boulder microporosity on Ryugu (see legend).}
\label{fig2}
\end{figure*}

\subsection{Constraining the Parent Body From Porosity}
\label{const}
The bulk porosity calculated for objects with a varying size, accretion time, and initial porosity (Fig. \ref{fig1}) shows that relatively large (i.e., slowly cooling and with a higher lithostatic pressure) and early accreted (i.e., $^{26}$Al-rich) planetesimals compact more efficiently. By contrast, small and late accreted bodies with relatively small pressures remain highly porous. Generally speaking, bodies on the bottom right of each panel on Fig. \ref{fig1} retain only a thin porous blanket, while those on the top left deviate negligibly from initially constant porosity profiles. While the value of $\phi_{bulk}$ does not provide information about the exact distribution of the porosity, it shows that intermediate conditions with a trade-off between size and accretion time should produce bodies with a high relative volume of material matching the microporosity of the boulders $\phi_{boulder}$.

Of importance for the accretion of Ryugu from the remnants of a larger object is the location of such material within the parent body. The distribution of the porosity is shown in Fig. \ref{fig2} by means of porosity profiles established after the cessation of compaction for different values of $\phi_{0}$ and compared with the Ryugu boulder microporosity. Profiles for reference radii of $1$, $5$, $10$, and $100$ km and several accretion times shown in Fig. \ref{fig2} represent different types of structures, which can have up to three layers and are defined by a straightforward comparison with $\phi_{boulder}$: $\phi>\phi_{boulder}$, $\phi \approx \phi_{boulder}$, and $\phi<\phi_{boulder}$. The porosity profile always has its maximum at the surface and decreases more or less steeply with an increasing depth reaching a range of minimum values between zero and $\phi_{0}$, depending on the planetesimal size and accretion time. Since $\phi_{0}=0.5$ falls within the range of $\phi_{boulder}$, two specific structures arise only for this initial porosity:
\begin{enumerate}
\item a surface layer with $\phi\approx \phi_{boulder}$ atop of a largely compacted interior (for $R>5$ km for any accretion time $t_{0}$ and for $R\leq 5$ km and a small $t_{0}$);
\item one single layer with $\phi \approx \phi_{boulder}$ (e.g., $R=1$ km for $t_{0}\geq 1$ Myr and $R=5$ km for $t_{0} \approx 5$ Myr).
\suspend{enumerate}
The structures occurring for $\phi_{0}>0.55$ are:
\resume{enumerate}
\item a surface layer with $\phi > \phi_{boulder}$, an intermediate layer with $\phi\approx \phi_{boulder}$, and a more (but not necessarily completely) compacted "core" area (for $R\gtrsim 10$ km for any $t_{0}$ and $R<10$ km and small $t_{0}$);
\item a surface layer with $\phi > \phi_{boulder}$ atop of a "core" area with $\phi\approx \phi_{boulder}$ (only for $R<10$ km and late accretion);
\item one layer with  $\phi > \phi_{boulder}$ (for $R\approx 1$ km with a late accretion and occurring only for $\phi_{0}\geq 0.7$).
\end{enumerate}

\begin{figure*}
\begin{minipage}[ht]{8.5cm}
\setlength{\fboxsep}{0mm}
\centerline{\includegraphics[trim = 10mm 0mm 10mm 0mm, width=8.5cm]{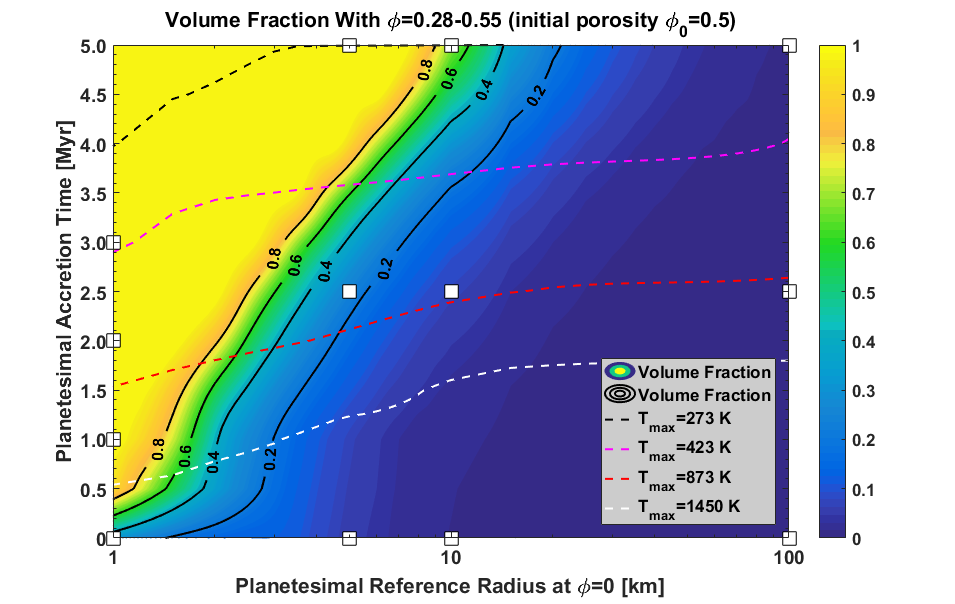}}
\end{minipage}
\begin{minipage}[ht]{8.5cm}
\setlength{\fboxsep}{0mm}
\centerline{\includegraphics[trim = 10mm 0mm 10mm 0mm, width=8.5cm]{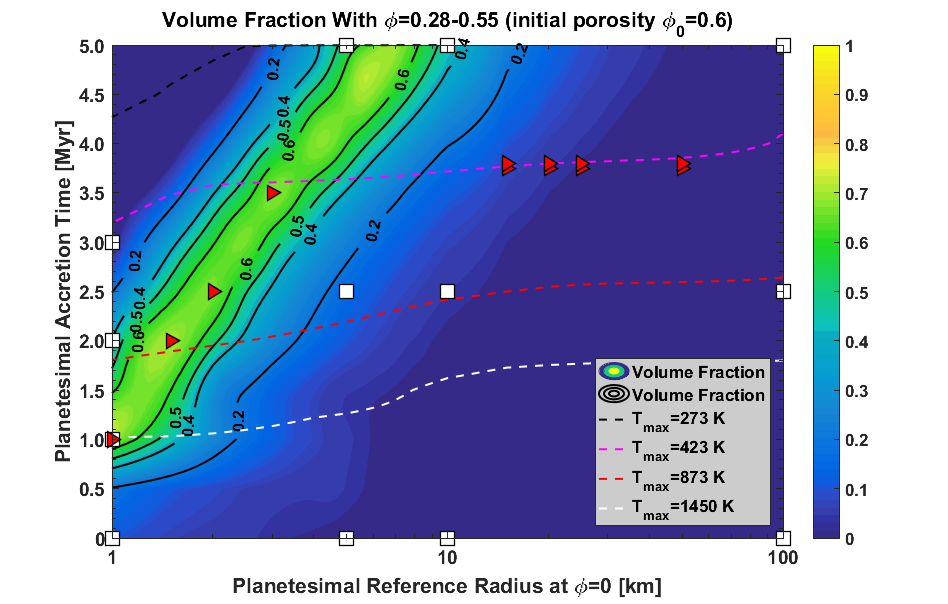}}
\end{minipage}
\\
\begin{minipage}[ht]{8.5cm}
\setlength{\fboxsep}{0mm}
\centerline{\includegraphics[trim = 10mm 0mm 10mm 0mm, width=8.5cm]{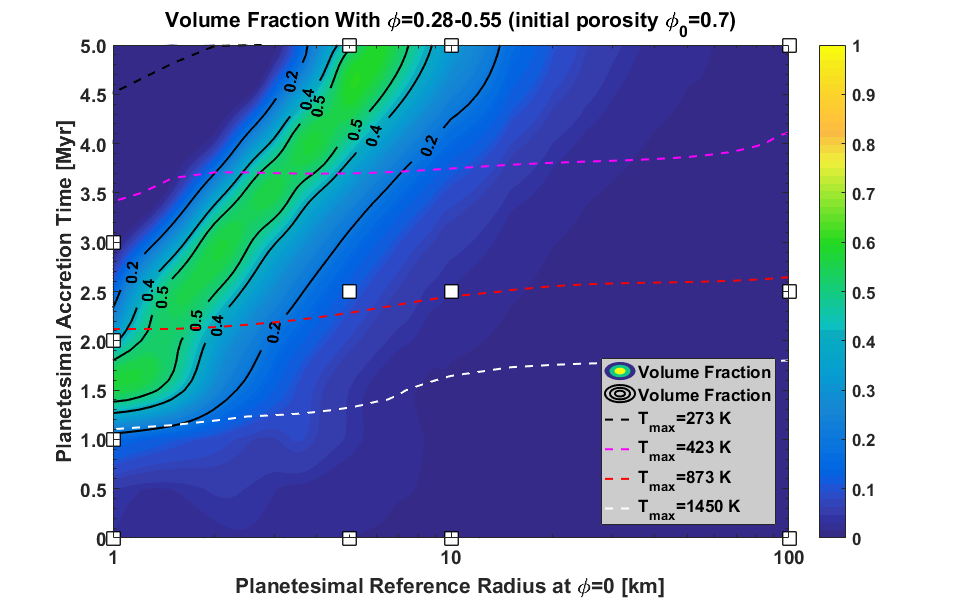}}
\end{minipage}
\begin{minipage}[ht]{8.5cm}
\setlength{\fboxsep}{0mm}
\centerline{\includegraphics[trim = 10mm 0mm 10mm 0mm, width=8.5cm]{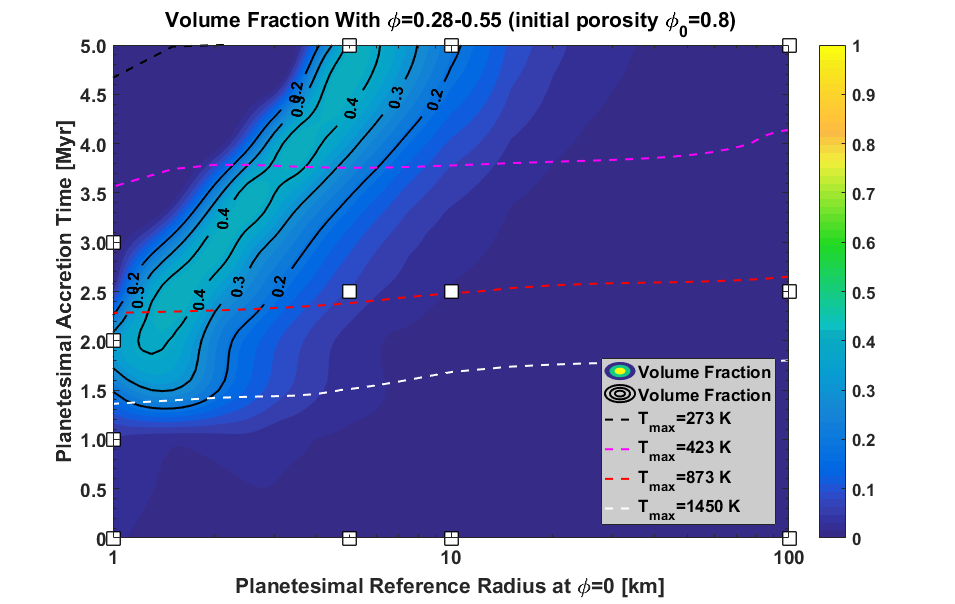}}
\end{minipage}
\caption{The volume fraction of the material with $\phi\approx\phi_{boulder}$, i.e., between $28$ \% and $55$ \%, as a function of the planetesimal reference radius (horizontal axes) and accretion time relative to the formation of CAIs (vertical axes). The panels correspond to different assumptions on the initial porosity $\phi_0$ of $50$ \% (top left), $60$ \% (top right), $70$ \% (bottom left), and $80$ \% (bottom right). Isolines for the maximum temperature $T_{max}$ are added for comparison: $273$ K (onset of aqueous alteration), $873$ K (serpentine dehydration), and $1450$ K (onset of silicate melting). White squares correspond to bodies addressed in Fig. \ref{fig2} while red triangles correspond to bodies shown in Fig. \ref{fig4}.}
\label{fig3}
\end{figure*}

The likelihood that a pair ($R$,$t_{0}$) identifies a parent body scales with the relative volume of the material with $\phi\approx \phi_{boulder}$, in particular, if materials from all depths contribute to each small object produced after the disruption of the original parent body \citep{Sugita2019}. Fig. \ref{fig3} shows the volume fraction defined as the ratio of the bulk volume with $\phi\approx\phi_{boulder}$ and the volume of the planetesimal:
\begin{eqnarray}
f_{boulder}:=V_{\phi\approx\phi_{boulder}}/V_{planetesimal}
\end{eqnarray}
within the ($R$,$t_{0}$) parameter space. The prominent area with $f_{boulder}>80$ \% of such material on the top left plot is simply owed to an initial porosity of $0.5$ that matches $\phi_{boulder}$ from the start on and to a weak compaction under the specific $(P,T)$ conditions in this area. For all higher initial porosities, diagonal field emerge (Fig. \ref{fig3}, top right, bottom left, and bottom right panels). The maximum value of $f_{boulder}$ is inversely proportional to $\phi_{0}$, such that for $\phi_{0}=0.8$, not more than $\approx 40$ \% of a planetesimal volume is in a favorable porosity range. As such, the consideration of $f_{boulder}$ confines a set of objects limited by a maximum radius of $\approx 10$ km and an accretion at $5$ Myr after CAIs, favoring an initial porosity that is close to $50$ \%. However, this is rather a technical effect of assuming an initial porosity within the range of the boulder porosity. Further factors, such as thermal conditions and alteration discussed in the following section and porosity estimates for small bodies \citep[e.g.,][]{Kiuchi2014,Paetzold2019} argue against this.

\subsection{Thermal Conditions and Alteration}
Another observation to be reproduced is aqueous alteration. Consideration of a sample chemical reaction of olivine and water to serpentine, talc, magnetite and hydrogen indicates quasi-instantaneous aqueous alteration on a geological time scale \citep{Jones2006,Neumann2020a,Wakita2011}. Thus, a first-order indicator and a lower limit for reproducing conditions for the formation of phyllosilicates is the melting temperature of water ice. It can be surpassed for a variety of accretion times and planetesimal sizes (Fig. \ref{fig3}, black dashed lines showing the maximum temperature of $273$ K), e.g., $t_{0} \lesssim 4$ Myr after CAIs for kilometer-sized objects (an effect of the rapid decay of $^{26}$Al), and also for a late accretion of larger bodies (an effect of their weaker cooling despite the lack of $^{26}$Al). As a rough upper limit on the thermal conditions inside the objects of interest, the production of silicate partial melt can be considered, occurring for $0.5\lesssim t_{0} \lesssim 1.8$ Myr after CAIs for varying values of $R$ and $\phi_{0}$ (Fig. \ref{fig3}, white dashed lines showing the maximum temperature of $1450$ K, i.e., the silicate solidus). Bodies to the lower right of the $T=1450$ K isoline produce even magma oceans. Under such conditions, the material is likely to dehydrate, such that the traces of aqueous alteration would be erased in the bulk of a planetesimal contradicting Ryugu’s observed composition.
\begin{figure*}
\begin{minipage}[ht]{5.6cm}
\setlength{\fboxsep}{0mm}
\centerline{\includegraphics[trim = 15mm 0mm 5mm 0mm, width=5.6cm]{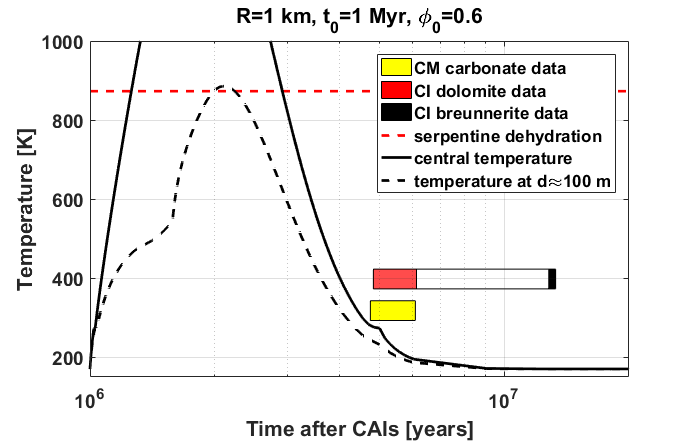}}
\end{minipage}
\begin{minipage}[ht]{5.6cm}
\setlength{\fboxsep}{0mm}
\centerline{\includegraphics[trim = 15mm 0mm 5mm 0mm, width=5.6cm]{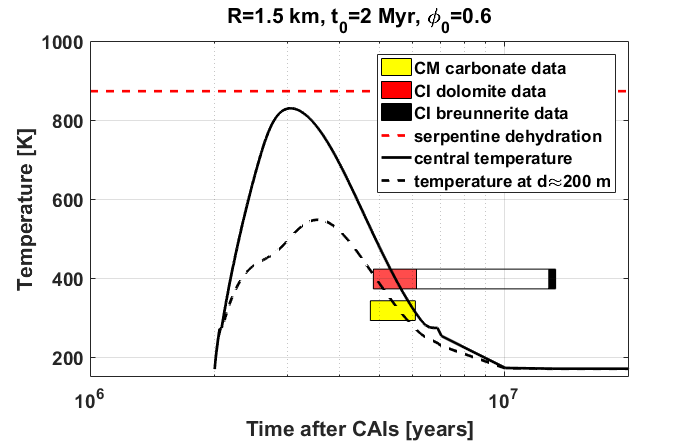}}
\end{minipage}
\begin{minipage}[ht]{5.6cm}
\setlength{\fboxsep}{0mm}
\centerline{\includegraphics[trim = 15mm 0mm 5mm 0mm, width=5.6cm]{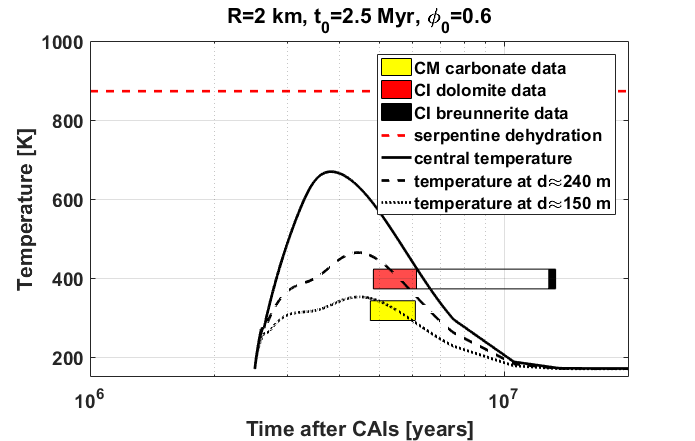}}
\end{minipage}
\\
[1ex]
\begin{minipage}[ht]{5.6cm}
\setlength{\fboxsep}{0mm}
\centerline{\includegraphics[trim = 15mm 0mm 5mm 0mm, width=5.6cm]{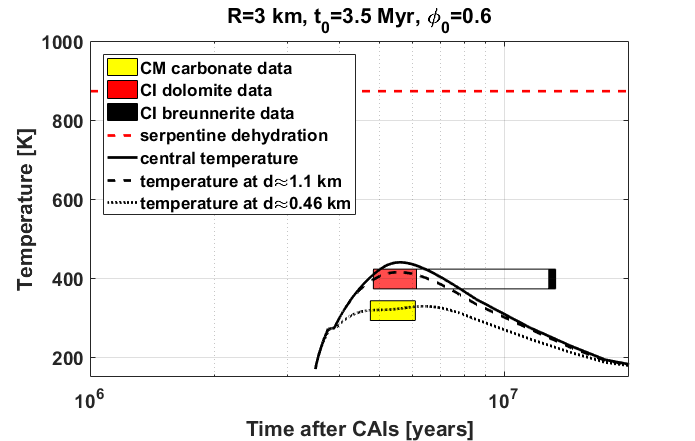}}
\end{minipage}
\begin{minipage}[ht]{5.6cm}
\setlength{\fboxsep}{0mm}
\centerline{\includegraphics[trim = 15mm 0mm 5mm 0mm, width=5.6cm]{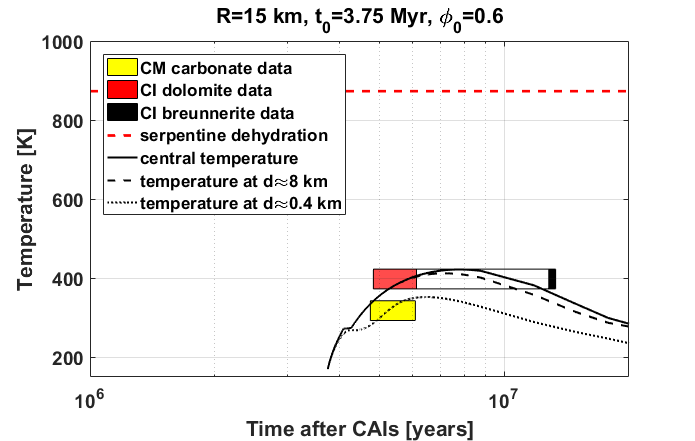}}
\end{minipage}
\begin{minipage}[ht]{5.6cm}
\setlength{\fboxsep}{0mm}
\centerline{\includegraphics[trim = 15mm 0mm 5mm 0mm, width=5.6cm]{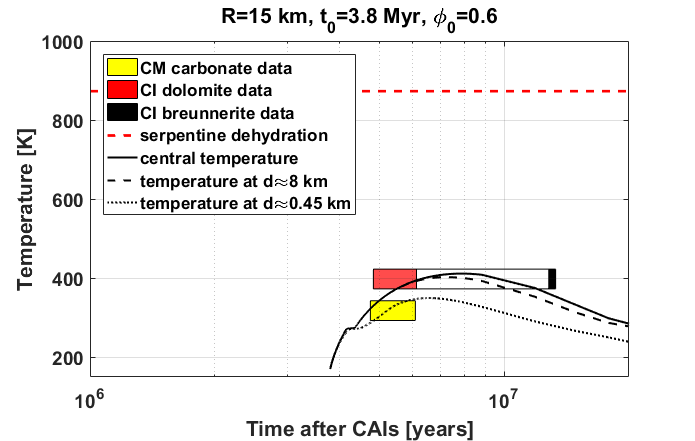}}
\end{minipage}
\\
[1ex]
\begin{minipage}[ht]{5.6cm}
\setlength{\fboxsep}{0mm}
\centerline{\includegraphics[trim = 10mm 0mm 5mm 0mm, width=5.6cm]{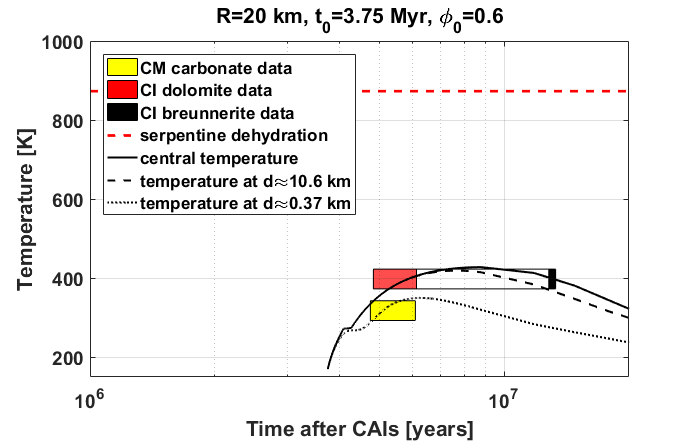}}
\end{minipage}
\begin{minipage}[ht]{5.6cm}
\setlength{\fboxsep}{0mm}
\centerline{\includegraphics[trim = 10mm 0mm 5mm 0mm, width=5.6cm]{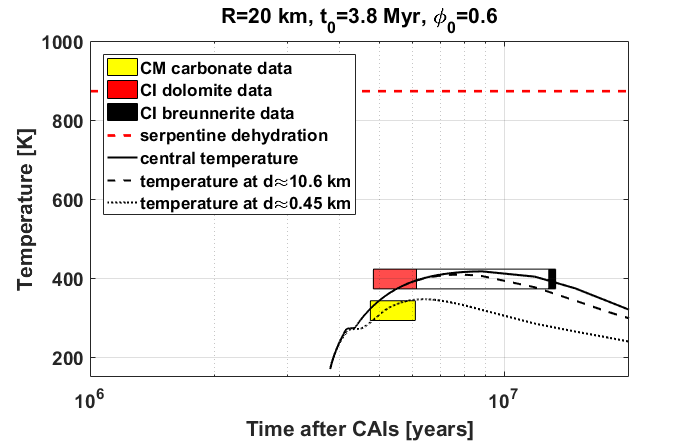}}
\end{minipage}
\begin{minipage}[ht]{5.6cm}
\setlength{\fboxsep}{0mm}
\centerline{\includegraphics[trim = 10mm 0mm 5mm 0mm, width=5.6cm]{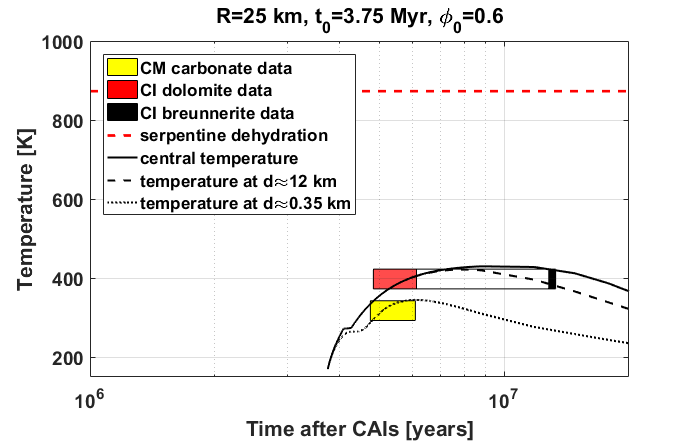}}
\end{minipage}
\\
[1ex]
\begin{minipage}[ht]{5.6cm}
\setlength{\fboxsep}{0mm}
\centerline{\includegraphics[trim = 10mm 0mm 5mm 0mm, width=5.6cm]{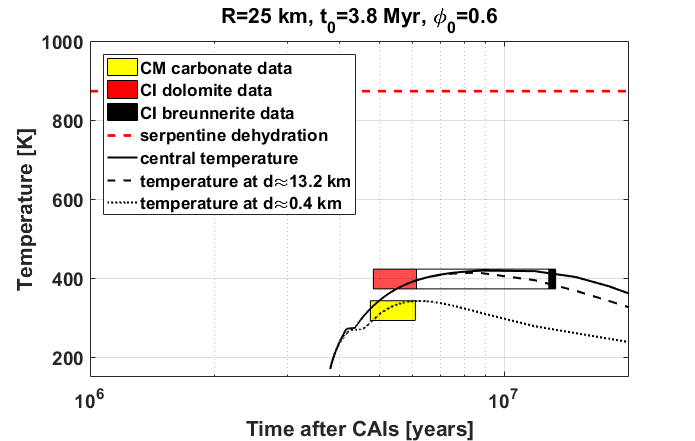}}
\end{minipage}
\begin{minipage}[ht]{5.6cm}
\setlength{\fboxsep}{0mm}
\centerline{\includegraphics[trim = 10mm 0mm 5mm 0mm, width=5.6cm]{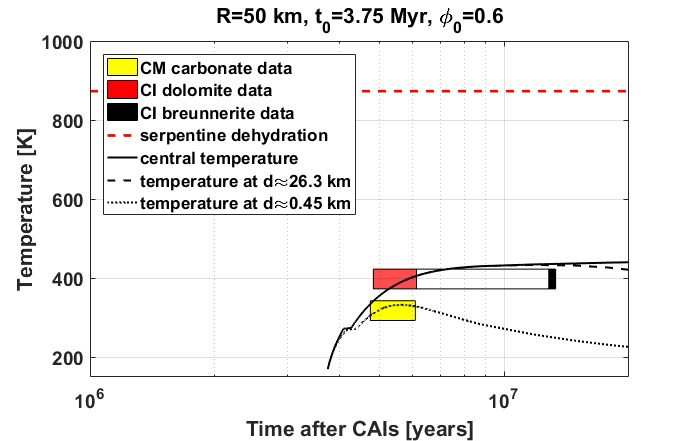}}
\end{minipage}
\begin{minipage}[ht]{5.6cm}
\setlength{\fboxsep}{0mm}
\centerline{\includegraphics[trim = 10mm 0mm 5mm 0mm, width=5.6cm]{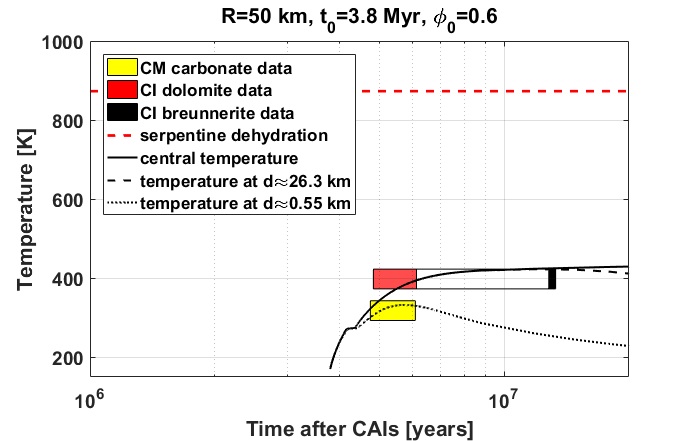}}
\end{minipage}
\caption{The evolution of the temperature at different depth $d$ compared with CI and CM carbonate formation times and temperatures. The temperature curves at the center and at depths that fit the data (where applicable) are shown. The empty black rectangle with the black bar represent the full time interval for the formation of breunnerite. Top row and left panel of middle row: Bodies with $\phi_{0}=0.6$ and $f_{boulder} \gtrsim 0.6$ (see Figs. \ref{fig1} and \ref{fig3}, top right panels). All other panels: Larger planetesimals with $\phi_{0}=0.6$ with accretion times of $3.75$ Myr and $3.8$ Myr after CAIs. Here, the CI dolomite and breunnerite formation times are either not fitted due to a relatively rapid cooling ($R=15$ km) or due to a longer heating phase ($R=50$ km), or they can be fitted quite well in some depth range ($R=20$ km and $R=25$ km).}
\label{fig4}
\end{figure*}
Thus, only the objects in the field approximately between the dashed black and red lines should have experienced thermal conditions that allowed for the production of hydrated material, and remained cool enough not to dry out. In particular, maximum temperatures obtained for a number of models satisfy the CI/CM alteration temperature range of $273-423$ K, while those obtained in the patch between red and white dashed lines indicate serpentine dehydration at $\approx 873$ K, consistent with the dehydration temperatures of the CI chondrite Y-86029 and CM chondrites Y-793321 and Jbilet Winselwan \citep{Velbel2019,King2018} shown to resemble closely Ryugu in their spectral appearance \citep{Sugita2019}. Despite some differences in the timing and curve shapes, the temperature trends shown in Fig. \ref{fig3} are alike for any $\phi_{0}$ considered. Differences in the temperature isoline shapes are consistent with progressively low values of the porosity-dependent thermal conductivity $k_{b}$ and high values of the actual radius $\overline{R}$ at high porosity, providing a stronger insulation of the interior and a weaker cooling of a planetesimal than at a low porosity.

It must be noted that any maximum temperature value shown in Fig. \ref{fig3} was attained at the center of a planetesimal and statements derived from $T_{max}$ refer, in the first place, to the material at the center and are not necessarily valid for the bulk of the object. For example, although for $\phi_{0}=0.5$ the fraction $f_{boulder}$ can be quite large with values of more than $0.8$ in a triagonal field (and, furthermore, $T_{max} \geq 273$ K in the most of the triagonal field, Fig. \ref{fig3}, top left panel), it is mostly not aqueously altered. For example, a body with $R=1$ km that accretes at $t_{0}=3$ Myr after CAIs will not be hydrated in its outer $300$ meters where the temperature does not surpass $273$ K at any time, such that the material that both satisfies the condition $\phi\approx \phi_{boulder}$ and is aqueously altered, has a volume fraction of only $\approx 0.4$. Therefore, with regard to both criteria, a low initial porosity of $<0.6$ still gives rise to a diagonal field of more probable parent bodies, similar to the cases with a higher $\phi_{0}$ (Fig. \ref{fig3}).

\subsection{Testing Connection to CI and CM Chondrites}
Testing a potential connection between Ryugu and the parent bodies of CI and CM meteorites, the evolution of the temperature evolution at different depths was compared with CI and CM carbonate formation ages and approximate temperatures. Results for selected models are presented in Fig. \ref{fig4}.

Carbonate formation occurred close to the peak of hydrothermal activity in CI and CM chondrites triggered by $^{26}$Al heating. Oxygen isotopic compositions of calcite and dolomite \citep{Clayton1984,Guo2007} and thermodynamic modeling \citep{Zolensky1989} yields $T\approx 293-343$ K and $T\approx 373-423$ K for CM and CI carbonates, respectively. As carbonates incorporate Mn (including the short-lived isotope $^{53}$Mn present at solar system formation) the decay system $^{53}$Mn-$^{53}$Cr can be used to constrain their formation age. 
Concerning Mn-Cr ages, we follow the calibration by \citet{Jilly2017}, who used a (U isotope corrected) U-Pb-Pb age of $4567.94\pm 0.31$ Myr for CAIs by \citet{Bouvier2011}, a  (U isotope corrected) U-Pb-Pb age of $4563.37\pm 0.25$ Myr for the Mn-Cr anchor D'Orbigny by \citet{Brennecka2012}, and a $^{53}$Mn/$^{55}$Mn ratio of $(3.54\pm 0.18)\times 10^{-6}$ by \citet{McKibbin2015}. This calibration of the $^{53}$Mn-$^{53}$Cr time scale results in Mn-Cr formation times ranging between $4.8$ and $6.1$ Myr after CAIs for CI and CM carbonates. In particular, the CM carbonate formation time interval $4.75-6.1$ Myr and the CI dolomite formation time interval of $4.83-6.14$ Myr are almost indisinguishable within error bars. In addition, the CI breunnerite formation occurred during an extended time interval, ranging between $6.1$ and $13.3$ Myr after CAIs \citep[see, e.g.,][and references therein]{Fujiya2013}.
\begin{table*}
\centering
\caption{Porosity in \% at the center and at the depths at which fits to carbonate formation data were established for bodies from Fig. \ref{fig4}. The value of $\approx 13$ \% established as lower bound in the interior of high-pressure planetesimals corresponds to a sluggish creep of olivine in the low-temperature regime.}
\centering
\begin{tabular}{lcccccccccccc}
\hline \\ [-1.7ex]
$(R,t_{0})$ & $(1,1)$ & $(1.5, 2)$ & $(2, 2.5)$ & $(3, 3.5)$ & $(15, 3.75)$ & $(15, 3.8)$ & $(20,3.75)$ & $(20,3.8)$ & $(25,3.75)$ & $(25,3.8)$ & $(50,3.75)$ & $(50,3.8)$ \\
\\ [-1.7ex]
\hline \hline
\\ [-1.4ex]
$\phi_{center}$ & $10$ & $27$ & $24$ & $28$ & $13$ & $13$ & $13$ & $13$ & $13$ & $13$ & $13$ & $13$ \\
 \\ [-1.4ex]
$\phi_{CI,br}$ & - & - & - & - & $13$ & $13$ & $13$ & $13$ & $13$ & $13$ & $13$ & $13$ \\
\\ [-1.4ex]
$\phi_{CI,dol}$ & - & $27$ & $54$ & $38$ & $13$ & $13$ & $13$ & $13$ & $13$ & $13$ & $13$ & $13$ \\
\\ [-1.4ex]
$\phi_{CM,car}$ & - & $54$ & $60$ & $54$ & $34$ & $33$ & $33$ & $30$ & $32$ & $30$ & $21$ & $16$ \\
[0.6ex]
\hline
\end{tabular}
\label{table2}
\end{table*}
For a comparison of Mn-Cr ages and thermal models of the parent body at various depths, the time-temperature curves should intersect the time-temperature regimes described above. The time-temperature regimes represented by boxes and optimal temperature curves as well as the associated depths are shown for selected models in Fig. \ref{fig4}.
Models show that such fits are possible only for formation times of $\lesssim 4$ Myr after CAIs. Therefore, on the one hand, several bodies from the favored parameter space with high values of $f_{boulder} \gtrsim 0.6$ that satisfy this condition are considered. On the other hand, also several larger objects that accreted before $4$ Myr after CAIs are included. In both cases, the analysis is restricted to $\phi_{0}=0.6$ for simplicity. Fig. \ref{fig4} compares temperature curves at different depths with the carbonate ages. For nearly all bodies shown, the CM carbonate and CI dolomite data can be fitted more or less successfully. However, for the first group, rapid cooling does not allow to fit the breunnerite data. In fact, this is not possible for $R<15$ km with any $t_{0}$. The porosities at fit depths shown in Table \ref{table2} indicate that in most cases the porosity at the fit depth surpasses the class average porosities of CI and CM chondrites. This was to be expected, since favored precursors of Ryugu discussed here have $f_{boulder} \gtrsim 0.6$. In addition, only for $R=1$ km dehydration of serpentine is indicated by the temperature evolution. Thus, a potential precursor of Ryugu with a radius of $1.5$ to $3$ km and an accretion time of $2$ to $3.5$ Myr, and $\phi_{0}=0.6$ could only be a parent body of CM or CI chondrites, if constrains on the partial dehydration of Ryugu's material, the porosity of CI and CM chondrites at temperature fit depths, and the breunnerite formation data are neglected. Furthermore, contrary to the general concept of a low maximum temperature of $\lesssim 423$ K throughout such a parent body, it would have heated up to $\geq 600-800$ K at its center, while remaining highly porous. Only the late-forming case of $R=3$ km, $t_{0}=3.5$ Myr after CAIs agrees with this concept, without being compatible with the late breunnerite formation time.

Shifting to the second group of larger and relatively late accreting bodies with $f_{boulder}<0.1$ allows for fitting the CI breunnerite data in addition to the CI dolomite in some cases. Here, it is required that a temperature curve crosses the CI dolomite data point, then  proceeds through the CI breunnerite rectangle without surpassing its maximum temperature, and then leaves it near the lower right corner. The latter condition corresponds to the youngest breunnerite age and no formation of this mineral after that. Such a condition cannot be fulfilled for an early accretion, since the temperature curves would either surpass the maximum temperature of $423$ K after crossing the dolomite data point, or stay within the temperature interval, but fall below the minimum temperature of $373$ K too early. For an accretion after $3.8$ Myr after CAIs, it is not possible to fit the dolomite formation data. Larger objects, in addition, do not satisfy the CI carbonate data, as it is shown for $R=50$ km. The CM carbonate data are fitted best at smaller depths of less than one km, while the CI carbonate data are fitted at larger depths, typically between a half-depth and the center of a planetesimal, where heat can be retained longer. In such cases, the temperature fit curve crosses the CI dolomite data range, indicating consecutive mineral formation, with dolomite precipitating during the prograde temperature evolution and breunnerite forming on both the prograde and the retrograde branch of the temperature curve at the same depth. None of the low-temperature bodies of the second group produce dehydrated material. Due to higher pressures relative to the first group, the microporosity is reduced to values between $0.13$ and $0.2$ throughout the interior except in thin shells of less than one kilometer. Average CI and CM class porosities are not reproduced at fit depths, but in thin outer shells. Under $(P,T)$ conditions occurring for different planetesimal radii, these thin shells have a higher relative thickness for a smaller reference radius, i.e., $R=15$ km. Naturally, bodies from the second group are rather bad candidates for being Ryugu's precursors due to a very small fraction $f_{boulder}$. However, some of them provide considerably better fits to the carbonate formation data including breunnerite and agree with the concept of a relatively homogeneous parent body that experienced a low maximum temperature of $\lesssim 423$ K throughout most of its interior, while retaining a substantial porosity (here, $>10$ \%). Although the microporosities of water-rich carbonaceous chondrites are reproduced only in very thin outer layers, bodies with radii of $15 \lesssim R \lesssim 25$ km are more suitable as parent bodies of CI and CM chondrites. While debris produced from these layers could contribute to Ryugu in a re-accretion event, the material involved would have experienced a maximum temperature far below that required for the dehydration of phyllosilicates, attributing partial dehydration to impact heating in this scenario.

\section{Summary and Conclusions}
\subsection{Planetesimal Bulk Microporosity}
Our modeling shows that icy bodies of different sizes and accretion times compact to considerably different values of the bulk microporosity (Fig. \ref{fig1}). The porosity reduction mechanism by hot pressing implemented here is likely the main compaction mechanism for the planetesimal parameter range considered. In this way, a complete lithification can be achieved in the deep interior of $^{26}$Al-rich bodies as small as a few km if a water ice rich primordial composition similar to those of CI and CM chondrites is assumed \citep{Neumann2014,Neumann2015}. Some mechanisms that could have influenced the porosity would require considerably higher pressures, such as cold pressing that can potentially reduce the porosity to $\approx 0.4$ \citep{Henke2012}. Other mechanisms, such as porosity change due to water reacting with dry silicates to less dense and more voluminous phyllosilicates while the water is consumed, and porosity change due to the opposite process of dehydration, are rather inefficient and were estimated to change the value of the porosity by $\lesssim 3$ \% \citep{Neumann2015}. Porosity reduction due to shock wave propagation during impacts \citep{Britt2002} that is not modeled here was apparently ineffective in the special case of Ryugu, given the high microporosity of its boulders.

\subsection{Parent Body Parameter Space}
For initial microporosities of $50$ \% to $80$ \%, Ryugu’s precursor could have evolved under different $(P , T)$ conditions over extended time periods of different length to reduce the microporosity to values below $55$ \% by compaction, while simultaneously avoiding it to drop below $28$ \%. While the porosity depth gradient follows the same general trend with a maximum equal to the initial value at the surface and a minimum at the center, the porosity depth distribution varies substantially. Consequently, models produce material with the boulder porosity in planetesimals of considerably different sizes and accretion times at different depths, i.e., in the deep interior (late accretion) or shallow layers (early accretion) of km-sized bodies and in shallow layers of $10-100$ km-sized bodies, but not in the upper unconsolidated blanket (except for $\phi_{0}=0.5$). Thereby, a volume fraction $f_{boulder}>0$ can be produced for any $\phi_{0}$ and for any pair $(R,t_{0})$ in some layer. However, its actual value varies greatly with $R$, $t_{0}$, and $\phi_{0}$, producing a maximum range for a diagonal field shown in Fig. \ref{fig3} (or a triagonal field for a special case of $\phi_{0} = 0.5$). Thermal boundary conditions of hydration (a lower bound) and dehydration (an upper bound) trim these areas further, i.e., conditions for the formation of phyllosilicates are given for $t_{0} \lesssim 4$ Myr for km-sized bodies and for late accreting larger bodies, while the post-dehydration process of melting of silicates occurs for $t_{0} \lesssim 1.8$ Myr for varying $R$ and $\phi_{0}$. Bodies with intermediate parameters experienced both hydration and partial dehydration, similar to Ryugu's inferred material and consistent with the dehydration temperatures of the Y-86029, Y-793321 and Jbilet Winselwan meteorites \citep{Velbel2019,King2018} as well as with metamorphic temperatures of Yamato-type chondrites \citep{King2019}.

\begin{figure*}
\setlength{\fboxsep}{0mm}
\centerline{\includegraphics[trim = 0mm 0mm 0mm 0mm, width=\textwidth]{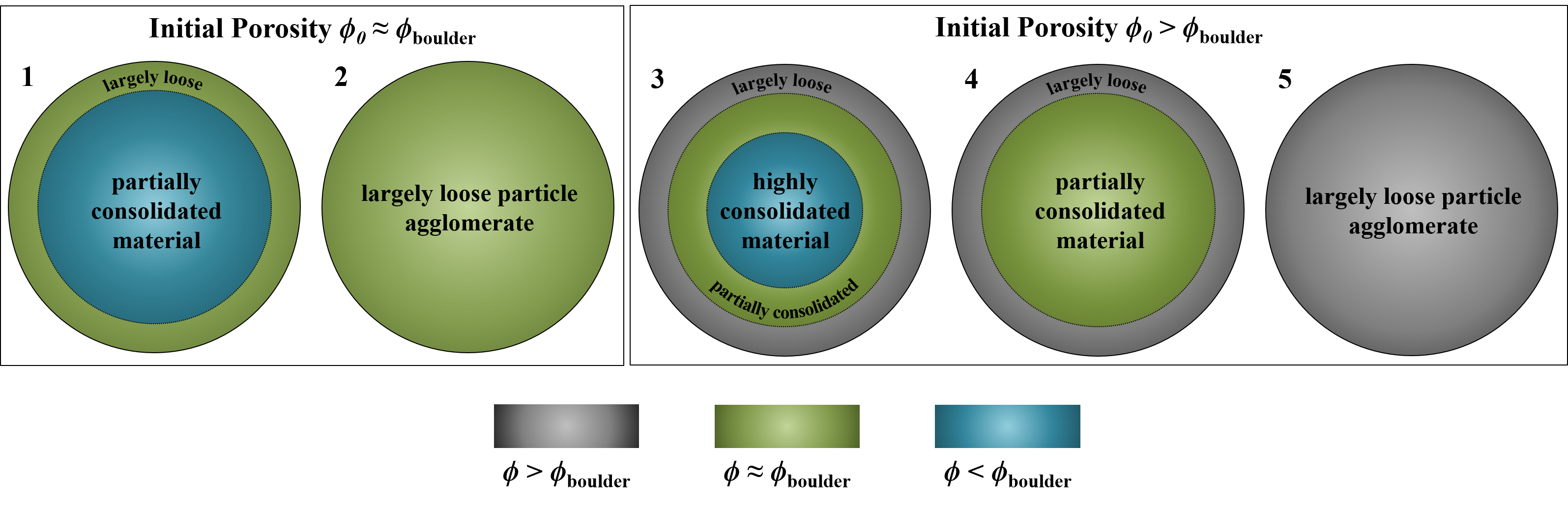}}
\caption{Structures obtained for the parent body of Ryugu described in section \ref{const}, numbered accordingly. The structures 1 and 2 are obtained for an initial porosity that falls in the range of the boulder porosity, while the structures 3-5 arise for higher values of the initial porosity. Structures 2 and 5 represent low-strength particle agglomerates that are likely to produce dust instead of boulders during an impact. Dust and high-density boulders would be produced from structure 1, dust and low-density boulders from structure 4, and dust as well as both low- and high-density boulders would be produced from structure 3.}
\label{fig5}
\end{figure*}

\subsection{Parent Body Structure}
Different parameters allow for the formation of objects with properties that are compatible with the observations and thus could represent the original parent body of Ryugu. The structures of these objects differ considerably from each other (Fig. \ref{fig5}). For an initial porosity within the range of the boulder porosity, a compacted interior below a loose surface layer, or a completely unlithified structure with $\phi \approx \phi_{boulder}$ are produced. The structures occurring for a higher initial porosity of $\phi_{0}>0.55$ can consist of up to three layers defined by a loosely packed particle aggregate with $\phi > \phi_{boulder}$, partially lithified material with $\phi\approx \phi_{boulder}$, and strongly compacted material with $\phi < \phi_{boulder}$. Clearly, those objects that started off with $\phi_{0} \approx \phi_{boulder}$ and compacted only negligibly, remain simply fluffy low-strength particle agglometares and would not be able to produce high-strength boulders. For all other above structures, boulder-like material is hidden at some depth below a fluffy envelope in consistence with a pressure-induced gradual lithification of the parent body material with depth. Our results imply formation of Ryugu from fragments of a highly porous parent body that experienced a low degree of lithification of initially porous material. Boulders might be formed during a break-up of such an asteroid from its consolidated interior, while outer layers that experienced negligible or no lithification would disperse to fine dust. A spectrum of boulder density and strength would occur according to the porosity profile of the parent body.

\subsection{Parent Body Evolution}
The evolution of the parent body derived here refines the formation scenario of Ryugu suggested by \citet{Okada2020}. After the accretion of silicate and water ice particles, the initial heating by $^{26}$Al, melting of water ice and hydration of dry silicates, our models produce a structure in which the porosity decreases continuously with depth but the planetesimal remains highly porous (Fig. \ref{fig2}). A stabilization of the porosity profile within a few million years after accretion implies that the break-up of the parent body might have occurred on a shorter time scale than its catastrophic collision timescale \citep{Wyatt2002}, suggesting a potentially early re-accretion of the rubble-pile material to form Ryugu. The low-temperature evolution of likely parent body candidates implies no dehydration of the material before the disruption of the parent body. Dehydration of serpentine at $\approx  873$ K is indicated for a very narrow set of parameters within the set of bodies with a high $f_{boulder}$. This result acts in favor of dehydration by impact heating. Either the original parent body was as small as $R\approx 1$ km and formed within $\approx 1.5$ Myr after CAIs, in which case it could be partially dehydrated, or the dehydration occurred during the catastrophic impacts. If dehydration is attributed to the impact process, then the formation time of the original parent body can be extended to at most $\approx 3.5$ Myr after CAIs (Fig. \ref{fig3}). Boulders of different densities can be attributed to re-accretion of partially consolidated material from different depths, but the dispersed dust from the surface layer could be incorporated as well. However, our models cannot exclude the origin of dense boulders from a dense impactor. Since none of the potential parent bodies satisfies $f_{boulder}=1$, the question of how can material from only specific layers with $\phi\approx\phi_{boulder}$ be exclusively re-accreted remains unanswered. While the upper loosely packed dust layer could disperse to dust during the collisions, exposing the boulder source layer to be fragmented by subsequent impacts, a less porous material from the central part would likely participate in the re-accretion. If so, and if more dense boulders originate from this part of the parent body, then they should be less porous than $\phi_{boulder}$. It should also be noted that bodies with a small $f_{boulder}$ would predominantly produce dense low-porosity boulders upon disruption producing a rubble-pile that would be considerably denser than Ryugu.

\subsection{Influence of Water on the Thermal History}
Spectral evidence for water action in the form of hydration of dry primordial minerals points, potentially, to water flow, although evidence for any sort of water flow on the parent body is lacking. Temperatures shown in Figs. \ref{fig3} and \ref{fig4} imply clearly melting of ice, hydration, dehydration, and evaporation of water for different parameter spaces. The most important impacts water could have on the thermal history are likely the latent heat buffering temperatures and fluid flow or hydrothermal convection allowing for more efficient heat transport. Both of these effects act towards a reduction of the internal temperatures and potentially expand the region in which porosity is stable. The latent heat of water ice melting is already included in the model and the temperature buffering effect is visible in the temperature evolution curves shown in Fig. \ref{fig4}. A fluid flow is not likely for the composition considered, since all or nearly all of the water that is produced after melting of water ice is consumed for the hydration of dry silicates. We still include an enhancement of the strain rate that a remaining free water volume fraction of $0.1$ would have, if water was not completely consumed during hydration. For such a small water fraction, an extensive water flow action is unlikely due to a low permeability. If this process were relevant for our study, then it would reduce internal temperatures and potentially shift the parameter space for favored parent bodies to earlier formation times. If hydration were considered as a not quasi-instantaneous process (e.g., occurring on a time scale of $^{26}$Al heating of $\approx 0.7$ Ma), then water flow or even water-rock differentiation may be of importance for the temperature evolution. However, consideration of the chemical reaction equation argues for fast hydration.

\subsection{Microporosity After Re-Accretion}
It is more likely that both compaction and aqueous alteration occurred in the original parent body. The occurrence of these processes in the re-accreted “secondary” parent body is indicated only if the disruption of the original parent body and the re-accretion event happened very early. If a re-accretion event occurred relatively late with a smaller concentration of $^{26}$Al, i.e., after $\approx 5$ Myr after CAIs, the secondary parent body should be definitely larger than several tens of kilometers for a microporosity change and consolidation to occur (but precursors favored here are smaller, see Fig. \ref{fig3}), if it existed in the first place. However, a catastrophic collision timescale of more than tens of Myr for the planetesimal size range considered here \citep{Wyatt2002} implies late disruption and re-accretion and, therefore, a microporosity change only on the original parent body.

\subsection{Collisional Lifetime}
A scenario with only a single disruption event could represent a challenge with respect to the collisional lifetime of small parent bodies favored by our study, i.e., the asteroid's mean lifetime before a catastrophic disruption occurs \citep[or, more precisely, the time for which the probability of survival is $1/e$, e.g.,][]{Durda1997,Nesvorny2017}. However, the collisional lifetime of a parent body with a radius of $\approx 3$ km is close to the age of the solar system \citep[][Fig. 14]{Bottke2005a}, meaning that a large fraction of such bodies will disrupt. This implies that such an object could survive for long enough to be destroyed in a single impact event, such that a part of its material re-accreted as the relatively young NEA Ryugu \citep[from $<10$ Ma to $\lesssim 1$ Ga, as indicated by surface age estimates by][]{Arakawa2020}, although this is not the most likely scenario. Another possibility is that of several disruption events with intermediate re-accreted parent bodies, which also occurs if the collisional lifetime is much shorter than the age of the solar system. In fact, impact models for the primordial main belt imply that most sub-100 km bodies are fragments of big collisions \citep{Bottke2005b}. If the collisional lifetime is small compared to the solar system age (but longer than the porosity evolution timescale of $\mathcal{O}(1)-\mathcal{O}(10)$ Ma), catastrophic disruption would occur several times, so Ryugu would be a result of several disruption and accumulation events. Since the thermal evolution is dominated by $^{26}$Al, it is reasonable to consider the microporosity evolution of the first generation of the parent body that are covered by our calculations. This would necessitate that subsequent disruption events do not change the microporosity to a notable degree. This is supported by impact models \citep{Bottke2005b} that show that a part of the material is not compacted during a collision.

\subsection{Relation to CI and CM Parent Bodies}
For the cases shown in Fig. \ref{fig4} and for bodies with a high fraction $f_{boulder}$ that accreted prior to $\approx 3.5$ Myr after CAIs, hydration of silicates occurs in the most part of the interior. The calculated heating of potential precursors with a high $f_{boulder}$ is consistent, in general, with the alteration temperatures of CI and CM chondrites \citep{Fujiya2012,Fujiya2013,Neumann2019c}. However, rapid cooling of those models that fit the CM carbonate and CI dolomite formation ages is inconsistent with CI breunnerite formation age, and their porosities mostly supercede those typical for CI and CM chondrites. Therefore, it is unlikely that Ryugu's parent body was simultaneously the parent body of one of these meteorite groups. Nevertheless, mineralogies that would appear similar to CI and CM chondrites in their spectral image are produced. More consistent with the meteorite data that suggest a relatively homogeneous parent asteroid with a maximum temperature of $\lesssim 423$ K are planetesimals with radii of $15$ km to $25$ km and an accretion time of $\approx 3.75-3.8$ Myr after CAIs. In particular, it is possible to fit both CI and CM carbonate formation data at different depths and manage the challenge provided by potentially young breunnerite formation ages in one parent asteroid. Here, objects in the range of $20$ km to $25$ km are preferred, since they satisfy carbonate formation conditions in their central areas. A challenge is represented by the average porosities of CI and CM chondrites of $34.9$ \% and $22.2$ \%, respectively. Porosity modeling indicates a stronger material consolidation at a higher temperature and pressure and a monotonous porosity decrease with depth, but higher CI carbonate formation temperature and higher CI porosities relative to CM chondrites conflict with that if both meteorite classes originated from the same parent body. A CM carbonate precipitation in a medium-temperature regime of $323-570$ K \citep{Verdier2017} could offer a way to resolve this contradiction. Although relatively large parent bodies derived for CI and CM classes are less favored as Ryugu's original parent body based on their small $f_{boulder}$, re-accretion of a small part from specific depth region of such a body provides another formation scenario for Ryugu, which would then be a rubble fragment of a CI or CM parent body. However, this rubble would originate from shallower depths than those where CI carbonate formation data were fitted (Table \ref{table2}) and a partial dehydration observed on Ryugu would be attributed to impact heating.

\subsection{Implications For Early Solar System Planetesimal Populations}
Lower average microporosities of carbonaceous chondrites than that of Ryugu imply that compaction was more pronounced on their parent bodies. This cannot be attributed to a potentially stronger heating, since peak temperatures inferred for CI and CM chondrites by multiple studies are $\approx 423$ K. Another option is the pressure effect on the evolution of the microporosity, that implies that parent bodies of hydrated carbonaceous chondrites were larger than Ryugu's original parent body. As shown further above, CI and CM meteorite data are easier to fit with a thermal evolution of a late-accreted larger object with a relatively low peak temperature and slow cooling. The results of our modeling suggest that Ryugu's parent body had a size of only several kilometers and accreted earlier than parent bodies of typical CI and CM chondrites. If Ryugu is representative for NEAs, then this would point to a different asteroid population - rubble piles formed after disruption of small, early accreted, ice-rich planetesimals that were heated close to or above the dehydration temperatures of phyllosilicates. Their material retained a relatively high porosity of $\phi \gtrsim \phi_{boulder}$ as a consequence of low lithostatic pressures. The Yamato-type carbonaceous chondrites pertrographically distinct from the CM class indicate decomposition of Mg-Fe-rich carbonates at $773-1073$ K and thermal metamorphism after an initial aqueous alteration \citep{King2019}. These meteorites could originate from the above population of small early accreted planetesimals and be representative for Ryugu. By contrast, parent bodies of CI and CM chondrites would represent another population of ice-rich planetesimals that formed $\approx 3$ Myr later and remained relatively cold with peak temperatures of $\approx 400$ K. However, being more massive, these bodies were able to compact more efficiently to average porosities of carbonaceous chondrites. A weak consolidation of Ryugu's parent body material would result in a low compressional strength of small meteoroids that were ejected during its disruption and could have reached the Earth. A lack of high-porosity samples with $\phi>0.5$ in the meteorite collection can be attributed to the breakup of highly porous low-strength meteoroids during atmospheric entry, as suggested by \citet{Grott2019}. A variety of structures and mechanical strength properties calculated for carbonaceous chondrite-like planetesimals (Figs. \ref{fig1} and \ref{fig2}) imply that C-type asteroids likely represent a spectrum of consolidation states and mechanical strengths, resulting from temperature and pressure conditions they experienced with respect to their size and accretion time.

\subsection{Relevance for Planetary Defense Endeavors}
Planetary defense concepts include targeted impacts onto potentially hazardous asteroids \citep{Holsapple2012,Jutzi2014}. Impacts onto porous targets differ substantially from those onto non-porous ones, for example in the terms of impact energy dissipation and the limits of impactor sufficiency. Another important aspect is their strength in general (depending, among other factors, on their porosity). The Earth-threatening asteroids, some of which are NEAs, are potentially highly porous. Porosity and strength estimates for these bodies are important for impact threat mitigation studies. Our model is capable of providing porosity structure estimates for asteroids based on only a few observational constraints.
\\

\textit{Acknowledgements: We thank the reviewer Carver Bierson and two anonymous reviewers for constructive comments and helpful suggestions, as well as Patrick Michel for a productive discussion on collisional lifetimes of main belt asteroids. WN and MT acknowledge support by Klaus Tschira foundation. MH was funded by Geo.X, the Research Network for Geosciences in Berlin and Potsdam, grant number: SO\_087\_GeoX.}

\bibliographystyle{cas-model2-names}

\bibliography{main}

\end{document}